\documentclass[10pt,paper]{JHEP3}
\usepackage{epsfig}
\usepackage{amsmath}

\def\zetaslash{\zeta\hskip-5.5pt/\hskip0.5pt}
\def\kslash{k\hskip-5.5pt/\hskip0.5pt}

\skip\footins = 1\bigskipamount plus 2pt minus 4pt

\newcommand{\tr}{\mathop{\rm tr}}

\title{4-point effective actions in open and closed superstring theory}

\author{Osvaldo Chand\'{\i}a \\
Dipartamento de Fisica, Universit\`a degli Studi di Padova and
Istituto Nazionale di Fisica Nucleare, 
Sezione di Padova, Italia\\
E-mail: \email{chandia@pd.infn.it}.}

\author{Ricardo Medina\\
Instituto de Ci{\^e}ncias, Universidade Federal de Itajub\'{a}\\
Itajub\'a, Minas Gerais, Brazil\\
E-mail: \email{rmedina@unifei.edu.br}.}

\abstract{Recently the effective action for the 4-point
functions in abelian open superstring theory has been derived, 
giving an explicit construction of the bosonic and fermionic terms
of this infinite $\alpha'$ series. 
In the present work we generalize this
result to the nonabelian case. We test our result, at 
${\alpha'}^3$ and ${\alpha'}^4$ order, with several existing
versions for these terms, finding agreement in most of the cases.
We also apply these ideas to derive the effective action for the 4-point
functions of the NS-NS sector of closed superstring theory, to all order
in
$\alpha'$.}

\preprint{DFPD 03/TH/38}

\begin{document}

\section{Introduction}

The subject of the string $\alpha'$ corrections to the Maxwell and Yang-Mills
lagrangians, in its bosonic as well as in its supersymmetric generalization,
began a long time ago. The first striking result consisted in the discovery
of the Born-Infeld lagrangian \cite{Born} as describing
the abelian effective theory 
of interacting open massless strings \cite{Fradkin}. This result contains the
infinite $\alpha'$ series of corrections to the Maxwell lagrangian, but it does
not include the derivative terms of the series. Since then, the nonabelian 
${\alpha'}^2 F^4$ terms \cite{Tseytlin, Gross1} the nonabelian
${\alpha'}^3 D^2 F^4$ terms \cite{Kitazawa}\footnote{
This work also included some partial information of the ${\alpha'}^3 F^5$
terms.} 
and the abelian ${\alpha'}^4 \partial^4 F^4$ terms \cite{Andreev}
 were determined,
among other things. All these results were derived using the $\sigma$-model or
the scattering amplitude approach. \\
In the last decade new interest arose in this subject since the discovery
that the low energy dynamics of open superstrings captured an 
equivalent description in terms of D-branes \cite{Witten}. Some time
afterwards, the supersymmetric generalization of the Born-Infeld lagrangian
was constructed 
\cite{Cederwall0, Cederwall0p, Bergshoeff0, Aganagic}. \\
Up to now, including the previous results in \cite{Fradkin} and
\cite{Aganagic}, infinite $\alpha'$ series for low energy effective actions 
have only been obtained in the abelian case \cite{Wyllard1, Wyllard2}. 
Besides the attempt of proposing a symmetrized trace prescription as a
nonabelian generalization of the Born-Infeld lagrangian \cite{Tseytlin1},
in the nonabelian case an analogue result
is not known  and all the
information that has been derived is strictly perturbative in $\alpha'$.
In this context, the scattering amplitude approach has been 
used to derive boson-boson, boson-fermion and fermion-fermion
terms of the effective action which are only quartic in the 
fields \cite{Bergshoeff1, Bilal}, since the general wisdom of scattering
amplitudes in string theory
has traditionally gone up to 4-point amplitudes \cite{Schwarz} only. Indeed,
a complete 5-point amplitude has only been derived 
for the open superstring in the past year \cite{Nos}\footnote{In
\cite{Kitazawa} a partial computation of this 5-point amplitude was done.}.
It has been used to confirm not only the nonabelian ${\alpha'}^3 D^2F^4$ terms,
but also the nonabelian ${\alpha'}^3 F^5$ terms \cite{Koerber1}, in which some 
controversy went on \cite{Kitazawa, Refolli, Koerber1}.\\
In the last three years, several successful methods arose as an alternative
way to derive the $\alpha'$ corrections to the Super Yang-Mills
(SYM) lagrangian. This includes: the determination of the 1-loop effective
action for $N=4$ SYM \cite{Refolli, Grasso}; the requirement of certain BPS
solutions to the deformed equations of motion \cite{Fosse, Koerber1,
Koerber2}; the supersymmetric deformations to the SYM and the Super
Maxwell lagrangians \cite{DeRoo2, DeRoo3} and the construction of
supersymmetric invariants by different superspace techniques 
\cite{Cederwall1, Cederwall2, Drummond}. Some of these methods were even used
up to ${\alpha'}^4$ level \cite{Koerber2, Drummond}, but in spite of their
success in reproducing the open superstring effective action, $all$ of them 
required in some moment 
some knowledge of the
superstring 4-point amplitudes, to fix some undetermined coefficients.\\
Recently a new type of infinite $\alpha'$ series contribution to the
effective action has been derived \cite{DeRoo1}. It
consists in the determination of the abelian supersymmetric terms which are
sensible to 4-point amplitudes, at any order in $\alpha'$.\\  
In the present work we generalize this previous result to the nonabelian
case. The infinite $\alpha'$ corrections to the SYM action basically consists
in a succession of nonabelian four field terms, all derived from the
correponding ones in  the ${\cal O}({\alpha'}^2)$ SYM invariant
\cite{Cederwall1, Cederwall2}  by taking appropriate derivatives to it, in
direct analogy to the construction in \cite{DeRoo1}. \\  
We also include in this work the 4-point effective action for the
NS-NS sector of closed superstring theory, which is also an infinite $\alpha'$
series result derived along the same lines of \cite{DeRoo1}. 
In the closed superstring case, the determination of higher $\alpha'$
corrections is much more involved than the corresponding ones in the open
superstring case. The first type of $\alpha'$ corrections found for closed
superstrings consisted in the
derivation of the ${\alpha'}^3 R^4$ terms in the gravitational sector 
of type
II  theories \cite{Gross1, Grisaru}. Subsequently, a similar
analysis was done for the heterotic superstring \cite{Kikuchi,Cai,Gross2}. It
has been noted that there are no ${\alpha'}^4$ corrections to the (tree level)
4-point effective action in the case of closed
superstrings \cite{Metsaev}\footnote{We thank R. Bentin for calling our
attention to this point.}. Some higher derivative  terms have also been studied
in \cite{Russo, Green:1999pu}. \\  
Our paper is organized as follows. In
section \ref{4-point} we deal with the
nonabelian generalization of \cite{DeRoo1}. We include a brief analysis about
the linear supersymmetry of the effective action, 
which is possible to be constructed in the abelian case \cite{DeRoo1},
but not in the nonabelian one, if only knowledge of 4-point amplitudes is to
be considered. We have also included ${\alpha'}^3$ and partial ${\alpha'}^4$ 
tests of
our action, together with the determination of the ${\cal O}({\alpha'}^5)$
terms,
as an application. We do not comment about non linear supersymmetry 
in this work.
In section \ref{closed} we apply the ideas of
\cite{DeRoo1}
to determine the effective action for the 4-point functions of the NS-NS
sector of closed superstring theory. We confirm that this tree level action
does not contain ${\cal O}({\alpha'}^4)$ terms \cite{Metsaev}, and we also
derive some higher order $\alpha'$ terms of it. Some final remarks
and conclusions are given in section \ref{remarks}.

\section{Nonabelian open superstring 4-point effective action}
\label{4-point}

\subsection{Review of the abelian effective action}
In \cite{DeRoo1} a tree level effective action was proposed
for the
4-point functions of the abelian open
superstring. In our conventions (see Appendix \ref{conventions}) it is 
given by\footnote{We use the notation 
$\partial_k \cdot \partial_j = 
\frac{\partial}{\partial x_k^{\mu}} \frac{\partial}{\partial x_{j \mu}}$.} 
\begin{eqnarray}
S^{\ abel.}_{\rm eff} & = & \int d^{10} x \ \left[ \
\left\{ \ -\frac{1}{4}
F_{\mu \nu} F^{\mu \nu} + \frac{i}{2} \
\bar{\psi} \gamma^{\mu} \partial_{\mu} \psi \ \right\}  \right. \nonumber
\\
 &  & - \ g^2 {\alpha'}^2 \
\int \int \int \int \ \left\{ \frac{}{}\prod_{j=1}^{4} d^{10} x_j \
\delta^{(10)}(x-x_j) \frac{}{} \right\} \nonumber \\
& &
{\cal G}(\partial_1 \cdot \partial_2 +
\partial_3 \cdot \partial_4 \ , \partial_1 \cdot \partial_4 +
\partial_2 \cdot \partial_3 \ , \partial_1 \cdot \partial_3 +
\partial_2 \cdot \partial_4)
\nonumber \\
 & & \times \{ \frac{}{}
F_{\mu \nu}(x_1)F^{\nu \rho}(x_2)
F_{\rho \sigma}(x_3)F^{\sigma \mu}(x_4)
  - \frac{1}{4} \ F_{\mu \nu}(x_1)F^{\mu \nu}(x_2)
F_{\rho \sigma}(x_3)F^{\rho \sigma}(x_4) \nonumber \\
 & &
+ \ 2 i \ \bar{\psi}(x_1)\gamma^{\nu} \partial_{\rho}\psi(x_2)
F_{\nu \mu}(x_3)F^{\mu \rho}(x_4)
- \ i \ \bar{\psi}(x_1)
\gamma^{\mu \nu \rho} {\partial}^{\sigma} \psi(x_2) F_{\mu \nu}(x_3)
F_{\rho \sigma}(x_4) \nonumber \\
 & & \left. -\frac{1}{3} \ \bar{\psi}(x_1)\gamma_{\mu}
{\partial}_{\nu} \psi(x_2) \bar{\psi}(x_3)\gamma^{\mu}
{\partial}^{\nu} \psi(x_4) \frac{}{} \} \ \right] \ ,
\label{Seff-abelian}
\end{eqnarray}
where the term in brackets, in the first line of (\ref{Seff-abelian}), is
the
$D=10$ Super Maxwell lagrangian and
\begin{eqnarray}
{\cal G}(s, t, u) = \frac{\Gamma(- \alpha' s)\Gamma(- \alpha' t)}
{\Gamma(1- \alpha' s - \alpha' t)} +
\frac{\Gamma(- \alpha' t)\Gamma(- \alpha' u)}
{\Gamma(1- \alpha' t - \alpha' u)} +
\frac{\Gamma(- \alpha' u)\Gamma(- \alpha' s)}
{\Gamma(1- \alpha' u - \alpha' s)} \ .
\label{G}
\end{eqnarray}
Note that ${\cal G}(s, t, u)$ is a symmetric function of $s$, $t$ and
$u$.\\
In \cite{DeRoo1} it was confirmed that
\begin{enumerate}
\item The 4-photon scattering amplitude derived from
$S^{\ abel.}_{\rm eff}$, in
(\ref{Seff-abelian}), reproduces the corresponding tree level amplitude
derived
in  abelian Open Superstring theory:
\begin{eqnarray}
{\cal A}^{(4)} = 16 \ i \ g^2 {\alpha'}^2 (2 \pi)^{10}
\delta^{(10)}(k_1+k_2+k_3+k_4) \ {\cal G}(s, t, u) \
K_{\rm 4b} \ ,
\label{4photons}
\end{eqnarray}
where
\begin{eqnarray}
s = - 2 k_1 \cdot k_2, \ \ \
t = - 2 k_1 \cdot k_4, \ \ \
u = - 2 k_1 \cdot k_3, \ \ \
\label{Mandelstam}
\end{eqnarray}
are the Mandelstam variables and
$K_{\rm 4b}$  is the 4-boson kinematic
factor, given in eq.(\ref{Kbosons}) of Appendix
\ref{SYM}. The Mandelstam variables satisfy the relation
\begin{eqnarray}
s+t+u=0 \ .
\label{zero}
\end{eqnarray}

\item The action is supersymmetric at every order in $\alpha'$.
\end{enumerate}
The main idea considered in \cite{DeRoo1} for proposing
(\ref{Seff-abelian}) as
the abelian 4-point effective action was basically to make the
correspondence
\begin{eqnarray}
k^{\mu}_j \rightarrow -i \frac{\partial}{\partial x^j_{\mu}} \ \ \ \ \ \ \
\
(j=1,2,3,4.) \ ,
\label{correspondence}
\end{eqnarray}
in a symmetrized way, in the momentum dependent factor ${\cal G}(s, t, u)$
that comes in the scattering amplitude (\ref{4photons}), and that
then goes in the effective action (\ref{Seff-abelian}).
This symmetrized prescription consists in first writing the Mandelstam
variables as
\begin{eqnarray}
s = -k_1 \cdot k_2 - k_3 \cdot k_4, \ \ \ \
t = -k_1 \cdot k_4 - k_2 \cdot k_3 \ \ \mbox{and} \ \
u = -k_1 \cdot k_3 - k_2 \cdot k_4,
\label{symmprescription}
\end{eqnarray}
which are on-shell equivalent to
(\ref{Mandelstam}), and then using the correspondence
(\ref{correspondence}) as indicated above. As a result,
${\cal G}(s, t, u)$ becomes
the differential
operator ${\cal G}(\partial_1 \cdot \partial_2 +
\partial_3 \cdot \partial_4 \ , \partial_1 \cdot \partial_4 +
\partial_2 \cdot \partial_3 \ , \partial_1 \cdot \partial_3 +
\partial_2 \cdot \partial_4)$
appearing in (\ref{Seff-abelian}), which is completely symmetric in the
space-time coordinates. This simplifies the proof of the
supersymmetry of (\ref{Seff-abelian}), to every order in $\alpha'$, as
was seen in \cite{DeRoo1}. We will comment more about the issue of
Supersymmetry  in subsection \ref{Supersymmetry}.

\subsection{The nonabelian formula}
\label{nonabelian}

We now present the main result of section \ref{4-point}, which is
our
nonabelian generalization of the effective action in
(\ref{Seff-abelian}):
\begin{eqnarray}
S^{\ nonabel.}_{\rm eff} = S_{\rm SYM}
+ \ S_{\rm \alpha' corr.}^{\ nonabel.}  \ ,
\label{sum}
\end{eqnarray}
where
\begin{eqnarray}
S_{\rm SYM} & = & \int \ d^{10} x \
\mbox{tr}
\left\{ \ -\frac{1}{4} F_{\mu \nu} F^{\mu \nu} +  \frac{i}{2} \ \bar{\psi}
\gamma^{\mu} D_{\mu} \psi \ \right\}
\label{SuperYangMills}
\end{eqnarray}
is the D=10 Super Yang-Mills action and
\begin{eqnarray}
S_{\rm \alpha' corr.}^{\ nonabel.} & = & - \frac{1}{2} g^2 {\alpha'}^2 \
\int d^{10}x \int \int \int \int \
\left\{ \frac{}{}\prod_{j=1}^{4} d^{10} x_j \
\delta^{(10)}(x-x_j) \frac{}{} \right\}
\nonumber \\
 & & \times \
{\cal G}^{\ a_1 a_2 a_3 a_4}_{\rm \alpha' corr.}
\left( \frac{}{} D_{1} \cdot D_{2} + D_{3} \cdot D_{4} \ ,
D_{1} \cdot D_{4} + D_{2} \cdot D_{3} \ ,
D_{1} \cdot D_{3} + D_{2} \cdot D_{4} \frac{}{} \right)
\nonumber \\
 & & \times \left[ \frac{}{}
F_{a_1 \mu \nu}(x_1) F_{a_2}^{\nu \rho}(x_2)
F_{a_3 \rho \sigma}(x_3)F_{a_4}^{\sigma \mu}(x_4)
  - \frac{1}{4} \ F_{a_1 \mu \nu}(x_1)F_{a_2}^{\mu \nu}(x_2)
F_{a_3 \rho \sigma}(x_3)F_{a_4}^{\rho \sigma}(x_4) \right.
\nonumber \\
 & &
+ \ 2i \ \bar{\psi}_{a_1}(x_1)\gamma^{\nu} D_{\rho}\psi(x_2)_{a_2}
F_{a_3 \nu \mu}(x_3)F_{a_4}^{\mu \rho}(x_4)
- \ i \ \bar{\psi}_{a_1}(x_1)
\gamma^{\mu \nu \rho} D^{\sigma} \psi(x_2)_{a_2} F_{a_3 \mu \nu}(x_3)
F_{a_4 \rho \sigma}(x_4) \nonumber \\
 & & \left. - \frac{1}{3} \ \bar{\psi}_{a_1}(x_1)\gamma_{\mu}
D_{\nu} \psi(x_2)_{a_2} \bar{\psi}_{a_3}(x_3)\gamma^{\mu}
D^{\nu} \psi(x_4)_{a_4} 
\frac{}{} \right]
\label{Seff-nonabelian}
\end{eqnarray}
represents the piece of the effective action that contains the
infinite $\alpha'$ corrections that come from superstring theory.\\
The function ${\cal G}^{\ a_1 a_2 a_3 a_4}_{\rm \alpha' corr.}$ appearing
in
(\ref{Seff-nonabelian}) is given by
\begin{eqnarray}
{\cal G}^{\ a_1 a_2 a_3 a_4}_{\rm \alpha' corr.}(s, t, u)
& = &
\left[ \ \frac{}{}
\{ \mbox{tr}(\lambda^{a_1} \lambda^{a_2} \lambda^{a_3} \lambda^{a_4}) +
\mbox{tr}(\lambda^{a_4} \lambda^{a_3} \lambda^{a_2} \lambda^{a_1}) \} \
f(s, t)
\right.\nonumber \\
& &
\ + \{ \mbox{tr}(\lambda^{a_1} \lambda^{a_3} \lambda^{a_2}
\lambda^{a_4}) +
\mbox{tr}(\lambda^{a_4} \lambda^{a_2} \lambda^{a_3} \lambda^{a_1}) \} \
f(t, u)
\nonumber \\
& &
\left. \ + \{ \mbox{tr}(\lambda^{a_1} \lambda^{a_2}
\lambda^{a_4} \lambda^{a_3}) +
\mbox{tr}(\lambda^{a_3} \lambda^{a_4} \lambda^{a_2} \lambda^{a_1}) \} \
f(u, s)  \frac{}{} \right]
\label{Gnonabelian}
\end{eqnarray}
where
\begin{eqnarray}
f(s, t) & = & \frac{\Gamma(- \alpha' s)\Gamma(- \alpha' t)}
{\Gamma(1 - \alpha' s - \alpha' t)} - \frac{1}{{\alpha'}^2 st} \ .
\label{f}
\end{eqnarray}
In (\ref{f}) we have subtracted the poles to the Gamma factors since
this last contribution has been considered separately in the SYM action in
(\ref{sum}).\\  
Note that, after substituting the symmetrized Mandelstam variables 
(\ref{symmprescription}),
${\cal G}^{\ a_1 a_2 a_3 a_4}_{\rm \alpha' corr.}(s, t, u)$ is symmetric 
in the pairs ($k_j, a_j)$. \\ 
A convenient way of writing the $\alpha'$ expansion of
${\cal G}^{\ a_1 a_2 a_3 a_4}_{\rm \alpha' corr.}$ will be commented in
section
\ref{alphaexpansion}, to be used afterwards in
(\ref{Seff-nonabelian}).\\ As shown in Appendix \ref{derivation}, the
action
in  (\ref{Seff-nonabelian}) indeed reproduces the 4-particle
scattering amplitudes (at tree level) of nonabelian Open Superstring
theory
\cite{Schwarz}: 4-boson, 2-boson/2-fermion and 4-fermion
amplitudes\footnote{Amplitudes involving an odd number of fermions are
trivially null.}:
\begin{eqnarray}
{\cal A}^{(4)} &
= & 8 \ i \ g^2 {\alpha'}^2 (2 \pi)^{10} \delta^{10}(k_1 + k_2 +
k_3 +k_4)\times \nonumber \\&&{}\times  \Biggl[ \frac{\Gamma(- \alpha' s)
 \Gamma(- \alpha' t)}{\Gamma(1- \alpha' s - \alpha' t)}
\{ \tr(\lambda^{a_1} \lambda^{a_2} \lambda^{a_3} \lambda^{a_4}) +
\tr(\lambda^{a_4} \lambda^{a_3} \lambda^{a_2} \lambda^{a_1}) \}
 \nonumber \\ & &
\hphantom{{}\times  \Biggl[}
+\frac{\Gamma(- \alpha' t) \Gamma(- \alpha' u)}{\Gamma(1- \alpha' t -
\alpha' u)} \{ \tr(\lambda^{a_1} \lambda^{a_3} \lambda^{a_2}
\lambda^{a_4}) +
\tr(\lambda^{a_4} \lambda^{a_2} \lambda^{a_3} \lambda^{a_1}) \}
 \nonumber \\ & &
\hphantom{{}\times  \Biggl[}
 + \frac{\Gamma(- \alpha' u) \Gamma(- \alpha' s)}{\Gamma(1-
\alpha' u - \alpha' s)} \{ \tr(\lambda^{a_1} \lambda^{a_2}
\lambda^{a_4} \lambda^{a_3}) +
\tr(\lambda^{a_3} \lambda^{a_4} \lambda^{a_2} \lambda^{a_1}) \}
 \Biggr] \cdot K^{(4)} \ ,
\label{A4}
\end{eqnarray}
where $K^{(4)}$ is the kinematic factor associated to the corresponding
scattering process.\\
Two immediate tests of (\ref{Seff-nonabelian}) are the following:
\begin{enumerate}
\item $Abelian \ limit$:\\
Using (\ref{zero}) we
see that in this limit (\ref{Seff-nonabelian}) leads directly to
(\ref{Seff-abelian}).

\item ${\alpha'}^2 \ terms$:\\
Using that the
momentum dependent factor in (\ref{f})  behaves as
$f(s,t) = -\pi^2/6 + {\cal O}(\alpha')$,
it can easily be seen that the first superstring correction to the Super
Yang-Mills action is given by
\begin{eqnarray}
S^{\ nonabel.}_{\rm (2)} & = & \frac{\
\pi^2}{2} g^2 {\alpha'}^2   \int \ d^{10} x \ \mbox{str} \left\{ \
\frac{}{} {F}_{\mu \nu}{F}^{\nu \rho}
{F}_{\rho \sigma}{F}^{\sigma \mu} - \frac{1}{4} \ F_{\mu \nu}F^{\mu
\nu}  F_{\rho \sigma}F^{\rho \sigma}  \right. \nonumber \\
 & & \left.  + \ 2i \ \bar{\psi}\gamma^{\nu} D_{\rho}\psi
F_{\nu \mu}F^{\mu \rho} - \ i \ \bar{\psi}
\gamma^{\mu \nu \rho} D^{\sigma} \psi F_{\mu \nu}
F_{\rho \sigma}
-\frac{1}{3} \ \bar{\psi}\gamma_{\mu}
D_{\nu} \psi \bar{\psi}\gamma^{\mu}
D^{\nu} \psi
\frac{}{} \right \} \ ,\nonumber \\
\label{S2}
\end{eqnarray}
where ``str'' means the symmetrized trace\cite{Tseytlin1},
calculated as
\begin{eqnarray}
\mbox{str}(\lambda^{a_1} \lambda^{a_2} \lambda^{a_3} \lambda^{a_4}) & = &
\frac{1}{6} \left\{ \frac{}{} \tr(\lambda^{a_1} \lambda^{a_2}
\lambda^{a_3}
\lambda^{a_4}) + \tr(\lambda^{a_1} \lambda^{a_4} \lambda^{a_3}
\lambda^{a_2}) +
\tr(\lambda^{a_1} \lambda^{a_4} \lambda^{a_2} \lambda^{a_3}) \right.
\nonumber\\
& & \left. + \tr(\lambda^{a_1} \lambda^{a_3} \lambda^{a_2} \lambda^{a_4})
+ \tr(\lambda^{a_1} \lambda^{a_3} \lambda^{a_4} \lambda^{a_2}) +
\tr(\lambda^{a_1} \lambda^{a_2} \lambda^{a_4} \lambda^{a_3}) \frac{}{}
\right\}
\ . \ \ \ \
\label{symmtrace}
\end{eqnarray}
The terms in 
(\ref{S2}) match completely with
the corresponding ones of
\cite{Bergshoeff1}\footnote{See eq. (4.1) of this reference.} and are
equivalent to the ones first derived in
\cite{Cederwall1,Cederwall2}\footnote{The authors of \cite{Bergshoeff1}
checked there that their ${\alpha'}^2$ quartic terms coincide, up to
on-shell
terms and total derivatives, with the ones of
\cite{Cederwall1,Cederwall2}.}.
\end{enumerate}
We end this subsection recalling that, since the action (\ref{Seff-nonabelian})
is a D=10 Super Yang-Mills extension,
the $\lambda^a$ $U(N)$ generators are all in the adjoint representation,
${(\lambda^a)}^{bc} = -i f^{abc \ }$, and therefore the six trace terms
in (\ref{Gnonabelian}), (\ref{A4}) and
(\ref{symmtrace}) satisfy  $\mbox{tr}(\lambda^{a} \lambda^{b} \lambda^{c}
\lambda^{d}) = \mbox{tr}(\lambda^{d} \lambda^{c} \lambda^{b} \lambda^{a})$,
reducing its number to only three.\\

\subsection{Linear Supersymmetry}
\label{Supersymmetry}

In the abelian case 
(\ref{Seff-abelian}) 
the linear supersymmetry of the
effective action was  proved at all orders in $\alpha'$ by arguing that
\begin{enumerate}
\item The ${\alpha'}^2$ order lagrangian, ${\cal L}_{(2)}^{abel.}$, is
supersymmetric.   
\item The higher $\alpha'$ lagrangians are automatically
supersymmetric since they are constructed by the action of the differential
operator  ${\cal G}(\partial_1 \cdot \partial_2 +
\partial_3 \cdot \partial_4 \ , \partial_1 \cdot \partial_4 +
\partial_2 \cdot \partial_3 \ , \partial_1 \cdot \partial_3 +
\partial_2 \cdot \partial_4)$ on the nonlocal version of 
${\cal L}_{(2)}^{abel.}$,
${\cal G}$ being symmetric in the
spacetime coordinates.
\end{enumerate}
Now, the difference between our nonabelian action (\ref{Seff-nonabelian})
with the abelian one (\ref{Seff-abelian}) is basically
the covariantization of the terms. In the nonabelian case some care must be
taken with the supersymmetry
since,  already at ${\alpha'}^2$ order, additional terms should be
considered \cite{Cederwall2}. The complete ${\cal L}_{(2)}^{non \ abel.}$
lagrangian is given, in our notation, by \cite{Cederwall2}
\begin{eqnarray}
{\cal L}_{(2)}^{non \ abel.} & = & \frac{\
\pi^2}{2} g^2 {\alpha'}^2  \mbox{str} \left\{ \
\frac{}{} {F}_{\mu \nu}{F}^{\nu \rho}
{F}_{\rho \sigma}{F}^{\sigma \mu} - \frac{1}{4} \ F_{\mu \nu}F^{\mu
\nu}  F_{\rho \sigma}F^{\rho \sigma}  \right. \nonumber \\
 & & + \ 2i \ \bar{\psi}\gamma^{\nu} D_{\rho}\psi
F_{\nu \mu}F^{\mu \rho} - \ i \ \bar{\psi}
\gamma^{\mu \nu \rho} D^{\sigma} \psi F_{\mu \nu}
F_{\rho \sigma}
-\frac{1}{3} \ \bar{\psi}\gamma_{\mu}
D_{\nu} \psi \bar{\psi}\gamma^{\mu}
D^{\nu} \psi \nonumber \\
 & & \left. - \frac{7}{60} \ g \ F^{\mu \nu}
\bar{\psi} \gamma_{\mu \nu \rho} \psi
\left \{  \bar{\psi},
\gamma^{\rho} \psi \right \}
+ \frac{1}{360} \ g \ F^{\mu \nu}
\bar{\psi}\gamma^{\rho \sigma \tau} \psi
\left \{  \bar{\psi},
\gamma_{\mu \nu \rho \sigma \tau} \psi \right \}
\frac{}{} \right \} .
\label{L2}
\end{eqnarray}
The additional terms in the third line of (\ref{L2}), which we will abbreviate
as $F(\psi \gamma \psi)^2$, are needed to 
form a nonabelian supersymmetric
invariant, and they are only sensible to 5 and 6-point amplitudes.
It is tempting to include them nonlocally in our nonabelian action
(\ref{Seff-nonabelian}) in order to have, at any order in $\alpha'$,
a completely supersymmetric invariant action. We explore this
possibility in the following lines concluding that this would only
work at ${\alpha'}^2$ order.\\  
The linear supersymmetry
transformations are given by
\begin{eqnarray}
\delta A_\mu \ = \ 2 \ \bar \epsilon\gamma_\mu\psi, \ \ \ \ \
\delta\psi \ = \ F_{\mu\nu}(\gamma^{\mu\nu}\epsilon) \ .
\label{susytransf}
\end{eqnarray}
When analyzing the variation of
(\ref{Seff-nonabelian}) under the supersymmetric transformations
(\ref{susytransf}), we may write this variation as a sum of two contributions:
\begin{enumerate}
\item The one that considers the variation of the nonlocal
lagrangian (which now includes the $F(\psi \gamma \psi)^2$ terms). 
\item The one that considers the variation of \\
${\cal G}^{\ a_1 a_2 a_3 a_4}_{\rm \alpha' corr.}
\left( \frac{}{} D_{1} \cdot D_{2} + D_{3} \cdot D_{4} \ ,
D_{1} \cdot D_{4} + D_{2} \cdot D_{3} \ ,
D_{1} \cdot D_{3} + D_{2} \cdot D_{4} \frac{}{} \right)$.
\end{enumerate}
In the first situation, the piece containing the
variation of only four field terms
will vanish if we consider the symmetry of the
${\cal G}^{\ a_1 a_2 a_3 a_4}_{\rm \alpha' corr.}$ operator, in 
direct analogy to the analysis done in \cite{DeRoo1} for the abelian case. 
But the piece containing the variation of the 
$F (\psi \gamma \psi)^2$ terms and the $\delta A_{\mu}$ variations from the
covariant derivative terms, will also cancel. 
It is known that their sum cancels in the local lagrangian case 
\cite{Cederwall2}, but it is possible to prove that in the nonlocal case they
cancel as well. This can be seen as follows:\\
The nonlocal variation of the $F (\psi \gamma \psi)^2$ terms can generically
be written as
\begin{eqnarray}
\epsilon F_1 F_4 \psi_2 \{\psi,\psi\}_3 
\label{var1}
\end{eqnarray}
and the corresponding generic nonlocal $\delta A_{\mu}$ variation (from the
covariant derivative terms) has the form 
\begin{eqnarray}
\epsilon F_1 F_2 \psi_3 \{\psi, \psi\}_4 \ .
\label{var2}
\end{eqnarray}
In (\ref{var1}) and (\ref{var2}) we use $F_1$, to denote $F_{a_1}(x_1)$, etc.
and the spacetime indexes have been omitted since for all kinds of spacetime
index contractions the argument will be the same. It is clear that both terms
will combine since we can rename $2\rightarrow 4, 3 \rightarrow 2,
4\rightarrow 3$ in the second term and then the resultant combination will 
vanish, as it happens in the local case. \\ 
So, what would only remain to see is that 
the variation coming from the 
$\delta {\cal G}^{\ a_1 a_2 a_3 a_4}_{\rm \alpha' corr.}$
contribution vanishes. Only at the lowest order 
in $\alpha'$ this is true, since that contribution in
${\cal G}^{\ a_1 a_2 a_3 a_4}_{\rm \alpha' corr.}$ is only a constant,
having a null variation.
As soon as higher $\alpha'$ contributions are taken into account,
this is no longer valid since these contributions include the 
$A_{\mu}$ variations that arise from the covariant derivatives. 
Then it would be expected that, from ${\alpha'}^3$ order on,
new terms in the effective action would be
needed to cancel these variations \cite{DeRoo2}. \\
Our conclusion is that, using 4-point amplitudes it is not possible to
construct and effective action which is
has supersymmetric invariants at every order in $\alpha'$, as was done in the
abelian  case \cite{DeRoo1}.

\subsection{$\alpha'$ expansion}
\label{alphaexpansion}

In this section we review the corresponding one of \cite{Bilal}.
The $\alpha'$ expansion of $f(s,t)$ in (\ref{f}) can be
determined to any desired order in $\alpha'$ using
the Taylor expansion for $\mbox{ln} \ \Gamma(1+z)$
\footnote{See  formula (10.44c) of \cite{Arfken}, for example.},
\begin{eqnarray}
\mbox{ln} \ \Gamma(1+z) = -\gamma z +
\sum_{k=2}^{\infty} (-1)^k \frac{\zeta(k)}{k} z^k \
\ \ \ \ \ \ (-1 < z \leq 1) \ ,
\label{Taylor}
\end{eqnarray}
as was remarked in \cite{DeRoo2}. In (\ref{Taylor})
$\gamma$ is the Euler-Mascheroni constant and $\zeta(k)$ is the Riemann
Zeta
function. This leads to:
\begin{eqnarray}
{\alpha'}^2 f(s,t) = \frac{1}{st} \left[
\frac{}{} \mbox{exp} \left\{ \sum_{k=2}^{\infty} \frac{\zeta(k)}{k}
{\alpha'}^k  (s^k+t^k-(s+t)^k) \right \} - 1 \frac{}{} \right] \ .
\label{formula1}
\end{eqnarray}
In \cite{Bilal} it was seen that the first terms of the $\alpha'$ series
of
this last expression  could be rearranged as\footnote{The minus signs of
the
terms containing odd powers of $\zeta(3)$ in eq. (2.7) of \cite{Bilal}
have been changed.}:
\begin{eqnarray}
{\alpha'}^2 f(s, t) & = & -\frac{\pi^2}{6} {\alpha'}^2
+\zeta(3) u \ {\alpha'}^3 +
\left[ - \frac{\pi^4}{180}(s^2+t^2+u^2) + \frac{\pi^4}{120} st \right]
{\alpha'}^4 \nonumber \\
 & & + \left[ -\frac{\pi^2}{6} \zeta(3) \ stu + \frac{1}{2} \zeta(5)
(s^2+t^2+u^2) u \right] {\alpha'}^5 \nonumber \\
 & & + \left[ -\frac{\pi^6}{2880}(s^2+t^2+u^2)^2 +
\frac{\pi^6}{3024}(s^2+t^2+u^2)(st +\frac{u^2}{2}) \right.
\nonumber \\
 & & \left. \ \ \ \ + \left( \frac{\pi^6}{3024}
+ \frac{\zeta(3)^2}{2} \right)
(stu) \ u \right] {\alpha'}^6 + {\cal O}({\alpha'}^7) \ ,
\label{expansionf}
\end{eqnarray}
where the relation (\ref{zero}) was taken into account.\\
At every order in
$\alpha'$, the coefficients of the series in (\ref{expansionf})
have been split into
two types of terms, according to their appearance in the expansion of
the abelian factor ${\cal G}$ in (\ref{G}), after including the expansions
of
$f(t,u)$ and $f(u,s)$:
the first ones, which are  symmetric in $(s,t,u)$ are the ones which will
be present in ${\cal G}$, while the second ones will cancel, upon
using condition (\ref{zero}). In the nonabelian factor ${\cal G}^{\ a_1
a_2 a_3
a_4}_{\rm \alpha' corr.}$ the first type of terms will contribute as
coefficients of
$\mbox{str}(\lambda^{a_1}\lambda^{a_2}\lambda^{a_3}\lambda^{a_4})$, while
the
second ones will contribute as coefficients of $f$-$f$ terms, in the
following
way:
\begin{eqnarray}
{\alpha'}^2 \ {\cal G}^{\ a_1 a_2 a_3 a_4}_{\rm \alpha' corr.}(s,t,u)= & &
\left\{ -\pi^2 \ \mbox{str}(\lambda^{a_1}\lambda^{a_2}
\lambda^{a_3}\lambda^{a_4}) \frac{}{} \right\} {\alpha'}^2
\nonumber \\
 & + & \left\{ - \zeta(3) \
\left[ f^{a_1 a_2 b} f^{a_3 a_4}_{\ \ \ \ \ b} \ u +
f^{a_1 a_3 b} f^{a_2 a_4}_{\ \ \ \ \ b} \ s \right]
\frac{}{} \right\} {\alpha'}^3 \nonumber \\
 & + &
\left\{ -\frac{\pi^4}{24} \ (s^2 + t^2 + u^2) \
\mbox{str}(\lambda^{a_1}\lambda^{a_2} \lambda^{a_3}\lambda^{a_4})
+ \frac{\pi^4}{360} \
\left[ \frac{}{} s(u-t) f^{a_1 a_2 b} f^{a_3 a_4}_{\ \ \ \ \ b}
\right. \right. \nonumber \\
& &  \left. \left.
\frac{}{} + u(s-t) f^{a_1 a_3 b} f^{a_2 a_4}_{\ \ \ \ \ b}
+ t(s-u) f^{a_1 a_4 b} f^{a_2 a_3}_{\ \ \ \ \ b} \frac{}{}
\right] \frac{}{} \right\} {\alpha'}^4 \nonumber \\
& + &  \left\{ \frac{}{} -\pi^2 \zeta(3) \ stu \ \mbox{str}
(\lambda^{a_1}\lambda^{a_2}\lambda^{a_3}\lambda^{a_4})
\right. \nonumber \\
& & \left. - \frac{1}{2} \zeta(5) \ (s^2+t^2+u^2) \
( f^{a_1 a_2 b} f^{a_3 a_4}_{\ \ \ \ \ b} \ u +
f^{a_1 a_3 b} f^{a_2 a_4}_{\ \ \ \ \ b} \ s ) \frac{}{}
\right\} {\alpha'}^5 \nonumber \\
& + & \left\{
-\frac{\pi^6}{480} \ (s^2+t^2+u^2)^2 \
\mbox{str} (\lambda^{a_1}\lambda^{a_2}\lambda^{a_3}
\lambda^{a_4})  \right. \nonumber \\
& & + \frac{\pi^6}{6048} \ (s^2+t^2+u^2) \
\left[ \frac{}{} s(u-t) f^{a_1 a_2 b} f^{a_3 a_4}_{\ \ \ \ \ b}
\right. \nonumber \\
& & \left.
+ u(s-t) f^{a_1 a_3 b} f^{a_2 a_4}_{\ \ \ \ \ b}
+ t(s-u) f^{a_1 a_4 b} f^{a_2 a_3}_{\ \ \ \ \ b} \frac{}{}
\right] \nonumber \\
& & \left. - \left(\frac{\pi^6}{3024} + \frac{\zeta(3)^2}{2} \right) \
(stu) \left[ \frac{}{}
f^{a_1 a_2 b} f^{a_3 a_4}_{\ \ \ \ \ b} \ u +
f^{a_1 a_3 b} f^{a_2 a_4}_{\ \ \ \ \ b} \ s \frac{}{}
\right] \frac{}{} \right\} {\alpha'}^6 \nonumber \\
& + & {\cal O}({\alpha'}^7) \ .
\label{expansionG}
\end{eqnarray}
In deriving the $f$-$f$ terms of this expansion, the (\ref{zero})
condition
together with the relations (\ref{T}) and (\ref{DeltaT}) were used.\\

\subsection{${\cal O}({\alpha'}^3)$ and ${\cal O}({\alpha'}^4)$ tests}
\label{O3O4tests}

The ${\cal O}({\alpha'}^2)$ test of formula (\ref{Seff-nonabelian})
was already considered in subsection \ref{nonabelian}.
In the present subsection we will make connection with
higher $\alpha'$ terms of the effective action
which are sensible to 4-point amplitudes and which
are already known in the literature.\\
Using integration by parts, equations of motion and the Jacobi
identity (\ref{Jacobi}), we have confirmed that the 
${\cal O}({\alpha'}^3)$ terms of our effective action (\ref{Seff-nonabelian}),
obtained using the 
\begin{eqnarray}
{\alpha'}^2 \ {\cal G}^{\ a_1 a_2 a_3 a_4}_{\rm (1)}(s,t,u)= & &
- \ \zeta(3) \
\left[ f^{a_1 a_2 b} f^{a_3 a_4}_{\ \ \ \ \ b} \ u +
f^{a_1 a_3 b} f^{a_2 a_4}_{\ \ \ \ \ b} \ s \right] \
{\alpha'}^3
\label{G3}
\end{eqnarray}
part of (\ref{expansionG}),
match completely with the corresponding bosonic and fermionic terms of
\cite{Bilal}.\\
In the case of the ${\cal O}({\alpha'}^4)$ terms we have only considered the
symmetrized ones, obtained from the
\begin{eqnarray}
{\alpha'}^2 \ {\cal G}^{\ a_1 a_2 a_3 a_4}_{\rm (2)}(s,t,u)= & &
-\frac{\pi^4}{24} \ (s^2 + t^2 + u^2) \
\mbox{str}(\lambda^{a_1}\lambda^{a_2} \lambda^{a_3}\lambda^{a_4}) \ 
{\alpha'}^4
\label{G4}
\end{eqnarray}
part of (\ref{expansionG}). These terms are the ones which  survive in the
abelian limit. Besides using integration by parts and equations of motion,
we have used the identities of Appendix \ref{conventions} and we have arrived
to the conclusion that only for the four fermion terms we have
agreement with \cite{Bilal}. Although our four boson and two
boson/two fermion terms disagree with the ones in \cite{Bilal}, we have found
some further 
agreement when comparing them with other existing versions in the literature,
as we will comment in the next lines.\\
We have checked that the abelian limit of four boson terms agree with the
$\partial^4 F^4$ ones of \cite{Andreev}. We also expect our terms to agree with
the symmetrized trace ones that would be obtained from
\cite{Koerber2}\footnote{The complete ${\cal O}({\alpha'}^4)$ terms
of \cite{Koerber2}, which also include the bosonic terms sensible to 5 and
6-point amplitudes, were recently confirmed
in \cite{Nagaoka}.} since in this reference agreement was also found with
\cite{Andreev} in the abelian limit.\\
In the case of the two boson/two fermion terms, we have checked that, in the
abelian limit, they
are equivalent to the corresponding ones  in eq. (3.2) of
\cite{DeRoo3}, except for a global relative sign between the $\gamma^{\mu}$
and the $\gamma^{\mu \nu \rho}$ terms. \\

\subsection{Higher order terms in the effective action}
\label{O5O6terms}

As an application of our nonabelian generalization (\ref{Seff-nonabelian})
we
now calculate the nonabelian ${\cal O}({\alpha'}^5)$ terms of the
effective
action which, as far as we know, have not been calculated before. Instead
of
using the symmetrized prescription for the Mandelstam variables
(\ref{symmprescription}), we now use 
\begin{eqnarray}
s \rightarrow  2 \ D_1 \cdot D_2 \ , \ \ \
t \rightarrow  2 \ D_1 \cdot D_4 \ \ \ \mbox{and} \ \ \
u \rightarrow  2 \ D_1 \cdot D_3 \ \ \ .
\label{uts}
\end{eqnarray}
The terms of the effective action derived using this prescription
will be on-shell equivalent to the ones in (\ref{Seff-nonabelian}).\\
Using (\ref{uts}) for the ${\cal O}({\alpha'}^5)$ terms of
(\ref{expansionG})
in (\ref{Seff-nonabelian}), we have
\begin{eqnarray}
S_{\rm (5)}^{\ nonabel.} & = & 2 \ g^2 \ {\alpha'}^5
\int d^{10}x \int \int \int \int \
\left\{ \frac{}{}\prod_{j=1}^{4} d^{10} x_j \
\delta^{(10)}(x-x_j) \frac{}{} \right\}
\nonumber \\
 & & \times \ \left[ \frac{}{}
2 \ \pi^2 \zeta(3) \ \mbox{str}
(\lambda^{a_1}\lambda^{a_2}\lambda^{a_3}\lambda^{a_4}) \
(D_1 \cdot D_2) (D_1 \cdot D_4) (D_1 \cdot D_3)
\right. \nonumber \\
& & \ \ \ \ \ + \ \zeta(5) \
\left\{ \frac{}{} f^{a_1 a_2 b} f^{a_3 a_4}_{\ \ \ \ \ b} \
(D_1 \cdot D_3) +
f^{a_1 a_3 b} f^{a_2 a_4}_{\ \ \ \ \ b} \
(D_1 \cdot D_2) \frac{}{} \right\}
\nonumber \\
 & & \left. \ \ \ \ \ \ \ \ \ \ \ \ \ \ \ \ \cdot
\left \{ \frac{}{} (D_1 \cdot D_2)^2 +
(D_1 \cdot D_4)^2 +
(D_1 \cdot D_3)^2 \frac{}{} \right \} \frac{}{}
\right] \nonumber \\
 & & \times \left[ \frac{}{}
F_{a_1 \mu \nu}(x_1) F_{a_2}^{\nu \rho}(x_2)
F_{a_3 \rho \sigma}(x_3)F_{a_4}^{\sigma \mu}(x_4)
  - \frac{1}{4} \ F_{a_1 \mu \nu}(x_1)F_{a_2}^{\mu \nu}(x_2)
F_{a_3 \rho \sigma}(x_3)F_{a_4}^{\rho \sigma}(x_4) \right.
\nonumber \\
 & &
+ \ 2i \ \bar{\psi}_{a_1}(x_1)\gamma^{\nu} D_{\rho}\psi(x_2)_{a_2}
F_{a_3 \nu \mu}(x_3)F_{a_4}^{\mu \rho}(x_4)
- \ i \ \bar{\psi}_{a_1}(x_1)
\gamma^{\mu \nu \rho} D^{\sigma} \psi(x_2)_{a_2} F_{a_3 \mu \nu}(x_3)
F_{a_4 \rho \sigma}(x_4) \nonumber \\
 & &  \left. -\frac{1}{3} \ \bar{\psi}_{a_1}(x_1)\gamma_{\mu}
D_{\nu} \psi(x_2)_{a_2} \bar{\psi}_{a_3}(x_3)\gamma^{\mu}
D^{\nu} \psi(x_4)_{a_4} 
\frac{}{} \right] \ .
\label{S51}
\end{eqnarray}
After performing the delta function integrations this leads to
\begin{eqnarray}
S_{\rm (5)}^{\ nonabel.} = &  & 4 \pi^2 \zeta(3) \ g^2 \ {\alpha'}^5
\int d^{10}x \ \mbox{str} \left[ \frac{}{}
D_{\lambda} D_{\tau} D_{\kappa} F_{\mu \nu}
D^{\lambda} F^{\nu \rho}
D^{\tau} F_{\rho \sigma}
D^{\kappa} F^{\sigma \mu} \right.
\nonumber \\
 & &
- \frac{1}{4} \
D_{\lambda} D_{\tau} D_{\kappa} F_{\mu \nu}
D^{\lambda} F^{\mu \nu}
D^{\tau} F_{\rho \sigma}
D^{\kappa} F^{\rho \sigma}
+ \ 2i \
D_{\lambda} D_{\tau} D_{\kappa} \bar{\psi}
\gamma^{\nu}
D^{\lambda} D_{\rho}\psi
D^{\tau} F_{\nu \mu}
D^{\kappa} F^{\mu \rho}
\nonumber \\
 & &  - \ i \
D_{\lambda} D_{\tau} D_{\kappa} \bar{\psi}
\gamma^{\mu \nu \rho}
D^{\lambda} D^{\sigma} \psi
D^{\tau} F_{\mu \nu}
D^{\kappa} F_{\rho \sigma} \nonumber \\
 & & \left.  -\frac{1}{3} \
D_{\lambda} D_{\tau} D_{\kappa} \bar{\psi}
\gamma_{\mu}
D^{\lambda} D_{\nu} \psi
D^{\tau} \bar{\psi}
\gamma^{\mu}
D^{\kappa} D^{\nu} 
\frac{}{} \right]
\nonumber \\
& + &  \ 2 \ \zeta(5) \ g^2 \ {\alpha'}^5 \
{\cal S}_{(5), f \mbox{-} f} \ ,
\label{S52}
\end{eqnarray}
where ${\cal S}_{(5), f \mbox{-} f}$ represents the term with structure
constants, written explicitly in Appendix \ref{S5}.\\
The abelian bosonic terms of (\ref{S52}) match completely
with the corresponding ones of \cite{DeRoo1}\footnote{Except for a
minus sign that we have taken into account in our formula \ref{S52}.}.
These terms were compared in \cite{DeRoo1} with the six-derivative ones of
\cite{Wyllard2}, finding agreement, up to field redefinitions.\\
A feature that was not present in the previous ${\cal O}({\alpha'}^k)$
terms
(with $k = 2, 3, 4$), and that
appears in (\ref{S52}) and in $all$ higher order $\alpha'$
terms, is
the fact that,  for the same $k$-th order in $\alpha'$,
besides $\zeta(k)$ multiplying a supersymmetric invariant, other
global coefficients are present multiplying other supersymmetric invariants. 
These coefficients will contain, as part
of
them, products of $\zeta(n)$, where $n$ is $odd$ at least once
(as, for example, the $\pi^2 \zeta(3) \sim \zeta(2)\zeta(3)$ coefficient
in
(\ref{S52})). So, it is expected that the ratio between these 
different global coefficients
is
not a fractional number. This fact has been used in
\cite{DeRoo2, DeRoo3, DeRoo1}, among other things,
to study the independent supersymmetric invariants that arise in the
effective
action, when considering the Noether method.\\
It is straight forward to construct the explicit expression of the
${\cal O}({\alpha'}^6)$ (or higher order) terms of the effective action,
using
the series in (\ref{expansionG}).

\section{Closed superstring 4-point effective action}
\label{closed}

\subsection{The general formula}

Let us review some basic facts about graviton scattering amplitudes in type
II
superstrings \cite{Polchinski2}. The tree level amplitude of 
four gravitons is given by
\cite{Schwarz}
\begin{eqnarray}
{\cal A}_{grav.}^{(4)} & = & 16\ i\ \kappa^2 e^{2\Phi} \ G(s,t, u) \ 
K_{grav.}^{(4)} \ ,
\label{amp}
\end{eqnarray}
where
\begin{eqnarray}
G(s,t,u) & = & \frac{1}{stu} \ \frac{\Gamma(1-\frac{1}{4}\alpha'
s) \ \Gamma(1-\frac{1}{4}\alpha' t) \ \Gamma(1-\frac{1}{4}\alpha'
u)}{\Gamma(1+\frac{1}{4}\alpha' s) \ \Gamma(1+\frac{1}{4}\alpha'
s) \ \Gamma(1+\frac{1}{4}\alpha' s)} \ .
\label{Gclosed}
\end{eqnarray}
In (\ref{amp}) $\kappa$ is the ten-dimensional gravitational 
coupling constant, $s, t, u$ are the same Mandelstam variables 
of (\ref{Mandelstam}) and the kinematical factor 
$K_{grav.}^{(4)}$, which is of eighth order in momenta, is given by
\begin{eqnarray}
K_{grav.}^{(4)} & = & 
t_{(8)}^{\mu_1 \nu_1 \mu_2 \nu_2 \mu_3 \nu_3 \mu_4\nu_4} \
t_{(8)}^{\lambda_1 \rho_1 \lambda_2 \rho_2 \lambda_3 \rho_3 \lambda_4 \rho_4}
\ \prod_{j=1}^4 
\zeta_{j\mu_j\lambda_j}k_{j\nu_j}k_{j\rho_j} \ .
\label{kin}
\end{eqnarray}
In (\ref{kin}), the $t_{(8)}$ tensor is exactly the same one
appearing in the four gluon scattering amplitude (\ref{Kbosons}) and
$\zeta_{\mu \nu} = \zeta_{\nu \mu}$ is the graviton polarization tensor.\\
We will see in the next subsection that the $\alpha'$ expansion of the Gamma
factor in (\ref{amp}) contains poles only in the leading term, in analogy to
the corresponding Gamma factor in the open superstring amplitude (\ref{A4}).
So, besides the leading contribution to the scattering amplitude, the rest of 
the $\alpha'$ terms contribute as 1-particle irreducible diagrams only,
being possible to apply the same technique of \cite{DeRoo1}.\\ 
It is very well known, that the first non zero $\alpha'$ contribution in the
effective action arises at order ${\alpha'}^3$ with the characteristic
coefficient $\zeta(3)$ and $R^4$ type of terms \cite{Gross1}. In particular
this means that, besides the leading D=10 Einstein-Hilbert term,
there are no ${\alpha'} R^2$ nor ${\alpha'}^2 R^3$ corrections in the
effective action. The ${\alpha'}^3 R^4$ term in the action which is
consistent with the ${\alpha'}^3$ terms of (\ref{amp}) has the form
\cite{Gross1, Polchinski2} 
\begin{eqnarray}
\frac{\zeta(3) \ {\alpha'}^3}{2^9 \cdot 4! \ \kappa^2} \int d^{10}x \ \sqrt{-g}
\ e^{-2\Phi} t_{(8)}^{\mu_1 \nu_1 \mu_2 \nu_2 \mu_3 \nu_3 \mu_4\nu_4} \
t_{(8)}^{\lambda_1 \rho_1 \lambda_2 \rho_2 \lambda_3 \rho_3 \lambda_4 \rho_4}
R_{\mu_1\nu_1\lambda_1\rho_1}
R_{\mu_2\nu_2\lambda_2\rho_2}
R_{\mu_3\nu_3\lambda_3\rho_3}
R_{\mu_4\nu_4\lambda_4\rho_4}. \nonumber \\
\label{sfork}
\end{eqnarray}
So, our proposal for the graviton 4-point functions in the effective action is
the following: 
\begin{multline}
S_{eff}^{\ grav.}  = {1\over 2 \kappa^2}\int d^{10}x \ \sqrt{-g} \ R \\
\begin{split}
& + {1 \over 2^{4} \cdot 4! \ \ \kappa^2} \int d^{10}x \
\sqrt{-g} \int  \left\{ \frac{}{} \prod_{j=1}^4 d^{10}x_j \  
\delta^{(10)}(x-x_j) \frac{}{} \right\} \ \hat G
(2 \nabla_{1} \cdot \nabla_{2} \ ,\ 
2 \nabla_{1} \cdot \nabla_{4} \ , \
2 \nabla_{1} \cdot \nabla_{3}) \\
& e^{-2\Phi} \
t_{(8)}^{\mu_1 \nu_1 \mu_2 \nu_2 \mu_3 \nu_3 \mu_4\nu_4}
t_{(8)}^{\lambda_1 \rho_1 \lambda_2 \rho_2 \lambda_3 \rho_3 \lambda_4 \rho_4} 
\ R_{\mu_1\nu_1\lambda_1\rho_1}(x_1) 
R_{\mu_2\nu_2\lambda_2\rho_2}(x_2)
R_{\mu_3\nu_3\lambda_3\rho_3}(x_3)
R_{\mu_4\nu_4\lambda_4\rho_4}(x_4) \ .
\end{split}
\label{Seff-R}
\end{multline}
In (\ref{Seff-R}) the $\hat G$ operator is defined  in terms of $G$
by means of 
\begin{eqnarray}
\hat G(s,t,u) & = & G(s, t, u) - \frac{1}{stu},
\label{Ggrav}
\end{eqnarray}
in analogy to (\ref{f}). In the next subsection it will become clear that the
${\alpha'}^3$ terms of (\ref{Seff-R}) match exactly with (\ref{sfork}). Note 
that, as there is no need to prove any 
supersymmetry\footnote{The complete action is indeed supersymmetric but,
at this level, we are
just dealing with the NS-NS sector which is completely bosonic.}, the
$\hat G$ operator does not need to be evaluated in a symmetrized way, as was
done in (\ref{Seff-abelian}) \cite{DeRoo1}. So, when using the prescription
(\ref{correspondence}), the Mandelstam variables have just been introduced as
in (\ref{Mandelstam}).\\ 
In analogy to the open superstring case (\ref{sum}),
(\ref{SuperYangMills}) and (\ref{Seff-nonabelian}),
in (\ref{Seff-R}) we have explicitly separated the leading contribution, given
by the D=10 Einstein-Hilbert term, from the rest of the $\alpha'$ correction
terms.\\
Strictly speaking, we would have to verify that the four graviton scattering
amplitude derived from (\ref{Seff-R}) is exactly expression (\ref{amp}), as
was done in \cite{DeRoo1}, and as we did in Appendix \ref{derivation} for
nonabelian open superstring . We have omitted this
proof since the procedure we would have to follow is directly the same one.\\ 
Now, what remains is to include the
rest of the NS-NS fields, namely, the Kalb-Ramond field $B_{\mu \nu}$ and the
dilaton $\Phi$.   The amplitude involving these states can be
calculated as before just noting that the dilaton polarization 
is a scalar and that the corresponding one for the $B_{\mu \nu}$
is an antisymmetric tensor, $\zeta_{\mu\nu}=-\zeta_{\nu\mu}$. 
In order to
derive the effective action for these fields, we would have to reconsider the
scattering amplitude approach for their ${\alpha'}^3$ terms, as was done for
the graviton. Fortunately, this has been done time ago \cite{Gross2}.
It was shown in \cite{Gross2} that this term is given exactly by (\ref{sfork})
by replacing the curvature tensor by the combination 
\begin{eqnarray}
{\bar R}_{\mu\nu}{}^{\lambda\rho}=R_{\mu\nu}{}^{\lambda\rho}+
 e^{-\Phi} \ \nabla_{[\mu}
H_{\nu]}{}^{\lambda\rho}-\delta_{[\mu}{}^{[\lambda}\nabla_{\nu]}
\nabla^{\rho]}\Phi \ , \label{Rbar}
\end{eqnarray}
where $H_{\mu \nu \lambda} =\partial_{[\mu} B_{\nu \lambda]}$. Therefore the
effective action for all the massless NS-NS states is given by
(\ref{Seff-closed}) by changing $R$ by $\bar R$. So our final expression for
the effective action NS-NS terms, sensible to 4-point amplitudes, is the
following:
\begin{multline}
S_{eff}^{\ \rm NS-NS}  = {1\over 2 \kappa^2}\int d^{10}x \ \sqrt{-g} \ 
\left( \frac{}{} R -\frac{1}{12} e^{-\Phi} H_{\mu \nu \lambda}H^{\mu \nu
\lambda} - \frac{1}{2} \ 
\partial_{\mu} \Phi \ \partial^{\mu} \Phi \frac{}{} \right) \\
\begin{split}
& + { 1 \over 2^{4} \cdot 4! \ \ \kappa^2} \int d^{10}x \
\sqrt{-g} \int  \left\{ \frac{}{} \prod_{j=1}^4 d^{10}x_j \  
\delta^{(10)}(x-x_j) \frac{}{} \right\} \ \hat G
(2 \nabla_{1} \cdot \nabla_{2} \ ,\ 
2 \nabla_{1} \cdot \nabla_{4} \ , \
2 \nabla_{1} \cdot \nabla_{3}) \\
& e^{-2\Phi} \ 
t_{(8)}^{\mu_1 \nu_1 \mu_2 \nu_2 \mu_3 \nu_3 \mu_4\nu_4}
t_{(8)}^{\lambda_1 \rho_1 \lambda_2 \rho_2 \lambda_3 \rho_3 \lambda_4 \rho_4} 
\ \bar{R}_{\mu_1\nu_1\lambda_1\rho_1}(x_1) 
\bar{R}_{\mu_2\nu_2\lambda_2\rho_2}(x_2)
\bar{R}_{\mu_3\nu_3\lambda_3\rho_3}(x_3)
\bar{R}_{\mu_4\nu_4\lambda_4\rho_4}(x_4) \ .
\end{split}
\label{Seff-closed}
\end{multline}

\subsection{Higher order terms in the effective action}

We will begin writing and explicit $\alpha'$ series for 
$\hat G(s, t, u)$, in analogy to what we did with
$f(s, t)$ in (\ref{expansionf}). For this purpose we 
first notice that
$G(s, t, u)$ in (\ref{Gclosed}) contains three factors of the form
$\Gamma(1-z) / \Gamma(1+z)$, that may be expanded using the Taylor series
(\ref{Taylor}), giving
\begin{eqnarray}
\ln\frac{\Gamma(1-z)}{\Gamma(1+z)}=2\gamma z
+2\sum_{n=1}^{\infty}\frac{\zeta(2n+1)}{2n+1} z^{2n+1} \ \ \ \ \ 
(|z| < 1 ) \ .
\label{formula2}
\end{eqnarray}
Using this result and condition (\ref{zero}), we then have that
\begin{eqnarray}
\hat G(s,t,u) & = & \frac{1}{stu} \left[ 
\exp\left\{ 2 \sum_{k=1}^{\infty}(\frac{\alpha'}{4})^{2k+1}\frac{\zeta(2k+1)}{
2k+1} (s^{2k+1}+t^{2k+1}+u^{2k+1}) \right\} \ - \ 1 \right] \ .
\label{Ghatexpansion}
\end{eqnarray}
This expression shows that the coefficients of the $\alpha'$ series of $\hat
G(s, t, u)$ are all given in terms of $odd$ valued $\zeta(n)$. Besides this,
it may be proved that $(s^{2k+1}+t^{2k+1}+u^{2k+1})$ is always divisible 
by $s t u$, when condition (\ref{zero}) is taken into account, so the series
in (\ref{Ghatexpansion}) has no poles. In fact, writing only the first
terms of the series, we have:
\begin{eqnarray}
\hat G(s, t, u) & = & 2 \ \zeta(3) \ (\frac{\alpha'}{4})^3  \ + \  
2 \ \zeta(5) \ (s^2+st+t^2) \ (\frac{\alpha'}{4})^5 \ 
+ \ 2 \ \zeta(3)^2 \ s t u \ (\frac{\alpha'}{4})^6  
\nonumber \\
& & + \ 2 \ \zeta(7) \ (s^4 + 2 \ t s^3 + 3 \ t^2 s^2 + 2 \ t^3 s + t^4) \
(\frac{\alpha'}{4})^7  \nonumber \\
& & + \ 4 \ \zeta(3)\zeta(5) \ stu(s^2+st+t^2)
\ (\frac{\alpha'}{4})^8 \ 
+ \ {\cal O}({\alpha'}^9) \ .
\label{formula3}
\end{eqnarray}
Let us write the first terms of the expansion in gravitational sector only.
The first term of the expansion is of order $(\alpha')^3$ and it 
is given by (\ref{sfork}) multiplied  $\zeta(3)$. Note that there is no 
term of order $(\alpha')^4$ \cite{Metsaev}. So the first new correction is at 
$(\alpha')^5$ and it is give by\footnote{We consider gravitational 
excitations only. Therefore, we use the combination $(\kappa e^{\Phi})$ as 
the gravitational constant $\kappa$ in the subsequent formulae.}

$$
\frac{\zeta(5) \ {\alpha'}^5}{2^{11} \cdot 4! \ \kappa^2} \int d^{10}x \ 
\sqrt{-g} 
\  t_{(8)}^{\mu_1 \nu_1 \mu_2 \nu_2 \mu_3 \nu_3 \mu_4\nu_4} \
t_{(8)}^{\lambda_1 \rho_1 \lambda_2 \rho_2 \lambda_3 \rho_3 \lambda_4 \rho_4}$$
$$
[ 2 \nabla_\sigma \nabla_\tau R_{\mu_1\nu_1\lambda_1\rho_1}
\nabla^\sigma \nabla^\tau R_{\mu_2\nu_2\lambda_2\rho_2}
R_{\mu_3\nu_3\lambda_3\rho_3}
R_{\mu_4\nu_4\lambda_4\rho_4} 
$$
\begin{equation}
+ \nabla_\sigma \nabla_\tau R_{\mu_1\nu_1\lambda_1\rho_1}
\nabla^\sigma R_{\mu_2\nu_2\lambda_2\rho_2} 
\nabla^\tau R_{\mu_3\nu_3\lambda_3\rho_3}
R_{\mu_4\nu_4\lambda_4\rho_4}].
\label{clo5}
\end{equation}
The next correction is at $(\alpha')^6$ and it is given by

$$
\frac{\zeta(3)^2 \ {\alpha'}^6}{2^{12} \cdot 4! \ \kappa^2} \int d^{10}x \ 
\sqrt{-g} 
\  t_{(8)}^{\mu_1 \nu_1 \mu_2 \nu_2 \mu_3 \nu_3 \mu_4\nu_4} \
t_{(8)}^{\lambda_1 \rho_1 \lambda_2 \rho_2 \lambda_3 \rho_3 \lambda_4 \rho_4}$$
\begin{equation}
[ \nabla_\sigma \nabla_\tau \nabla_\omega R_{\mu_1\nu_1\lambda_1\rho_1}
\nabla^\sigma R_{\mu_2\nu_2\lambda_2\rho_2} 
\nabla^\tau R_{\mu_3\nu_3\lambda_3\rho_3}
\nabla^\omega R_{\mu_4\nu_4\lambda_4\rho_4}].
\label{clo6}
\end{equation}
Finally we write the order $(\alpha')^7$

$$
\frac{\zeta(7) \ {\alpha'}^7}{2^{13} \cdot 4! \ \kappa^2} \int d^{10}x \ 
\sqrt{-g} 
\  t_{(8)}^{\mu_1 \nu_1 \mu_2 \nu_2 \mu_3 \nu_3 \mu_4\nu_4} \
t_{(8)}^{\lambda_1 \rho_1 \lambda_2 \rho_2 \lambda_3 \rho_3 \lambda_4 \rho_4}$$
$$
[ 2 \nabla_\sigma \nabla_\tau \nabla_\omega \nabla_\kappa
R_{\mu_1\nu_1\lambda_1\rho_1}
\nabla^\sigma \nabla^\tau \nabla^\omega \nabla^\kappa
R_{\mu_2\nu_2\lambda_2\rho_2}
R_{\mu_3\nu_3\lambda_3\rho_3}
R_{\mu_4\nu_4\lambda_4\rho_4} 
$$
$$
+4 \nabla_\sigma \nabla_\tau \nabla_\omega \nabla_\kappa
R_{\mu_1\nu_1\lambda_1\rho_1}
\nabla^\sigma \nabla^\tau \nabla^\omega 
R_{\mu_2\nu_2\lambda_2\rho_2}
\nabla^\kappa R_{\mu_3\nu_3\lambda_3\rho_3}
R_{\mu_4\nu_4\lambda_4\rho_4} 
$$
\begin{equation}
+ 3 \nabla_\sigma \nabla_\tau \nabla_\kappa R_{\mu_1\nu_1\lambda_1\rho_1}
\nabla^\sigma \nabla^\tau R_{\mu_2\nu_2\lambda_2\rho_2} 
\nabla^\omega  \nabla^\kappa R_{\mu_3\nu_3\lambda_3\rho_3}
R_{\mu_4\nu_4\lambda_4\rho_4}].
\label{clo7}
\end{equation}
It would be interesting to determine the supersymmetric extension of 
(\ref{Seff-closed}). In order to do this, it is necessary to compute 
scattering amplitudes involving up to 16 fermions. 
It will be also necessary to 
include Ramond-Ramond fields. It is much more 
convenient to use the constraints imposed by the corresponding supersymmetry 
by following \cite{Green:1998by} 
(recall there are two possible supersymmetries for the same NS-NS massless
fields, type IIA and type IIB). \\
The calculation we have done is at tree-level for the string. It is known that
the term of the effective action $\zeta(3) R^4$ 
receives corrections beyond one-loop 
and even non-perturbative ones \cite{Green:1997tv}. 
Namely, for both type II superstrings,  
this term receives corrections at tree and at one-loop level in 
string perturbation theory. 
For the type IIB superstring, this term also receives non-perturbative 
corrections. It is also known that the term $\zeta(5) \nabla^4 R^4$ receives 
perturbative corrections at tree-level, as we see in (\ref{clo5}), 
and at two-loops and non-perturbative ones for the type IIB
superstring \cite{Green:1999pu}. There has been investigations to determine 
the perturbative and non-perturbative structure of some higher derivative 
terms \cite{Russo}. It would be interesting to investigate if 
the remaining terms in  (\ref{Seff-closed}) have similar properties
\footnote{We thank to Pierre Vanhove for useful comments about these 
properties.}.

\section{Final remarks and conclusions}
\label{remarks}

We have derived two new results for effective
actions in superstring theory. Both of them consist
in explicit expressions for the infinite $\alpha'$ series and they are based 
in a recent derivation of the effective action for the 4-point functions in
abelian open superstring theory \cite{DeRoo1}.\\
Our first result consists in the nonabelian generalization of \cite{DeRoo1}.
We have tested our nonabelian action (\ref{Seff-nonabelian}), at ${\alpha'}^3$
and ${\alpha'}^4$ order, with the corresponding expressions calculated by 
other methods \cite{Andreev, Koerber2, Bilal, DeRoo3}, which are
different to the scattering  amplitude approach. In most of the cases we have
found agreement with the previous results.\\ 
Our second result consists in the effective action for the 4-point functions   
in the NS-NS sector of closed superstring theory. In the context of closed
superstring effective actions, we have reproduced the known results for 
terms at  ${\cal O}({\alpha'}^3)$ \cite{Gross1, Gross2} and 
explicitly found the tensor structure of some 
higher derivative terms discussed in \cite{Russo, Green:1999pu}. 
Our action
(\ref{Seff-closed}) also confirms the non existence of 
${\cal O}({\alpha'}^4)$ terms which are sensible to (tree level) 
4-point amplitudes \cite{Metsaev}.\\
We would like to stress
that the only Ansatz needed to compute our on-shell effective actions,
sensible to 4-point amplitudes, at
any desired order in $\alpha'$, is the one done in (\ref{Seff-nonabelian})
and (\ref{Seff-closed}). This should be contrasted with the previous 
traditional method of going, order by order in
$\alpha'$, proposing the most general structure of the four field terms, with
coefficients to be determined by consistency with superstring scattering
amplitudes (see \cite{Tseytlin1, Bilal}, for example).\\
We have used our effective actions (\ref{Seff-nonabelian}) and
(\ref{Seff-closed}) to derive higher order $\alpha'$ corrections, as an
application. These ${\cal O}({\alpha'}^k)$ terms are by no means a
complete list, in the sense that other terms which are sensible to higher
n-point scattering amplitudes ($n > 4$) should also be present at the same
${\alpha'}^k$ order.\\
As a final remark we would like to comment on the 
$\zeta(n)$ dependence in the $\alpha'$ series of the Gamma factors 
which appear in the 4-point amplitudes (\ref{A4}) and (\ref{amp}). It was 
used in \cite{DeRoo2} that the Taylor series for 
$\mbox{ln} \ \Gamma (1+z)$, eq.
(\ref{Taylor}), could make manifest the $\zeta(n)$ dependence 
of the Gamma factor for the open superstring (see eq. (\ref{formula1})). 
It is clear then, except for the zero $\alpha'$ order and the 
low order null coefficients, that 
the rest of the coefficients of the $\alpha'$ series
are all composed of $\zeta(n)$ factors \cite{DeRoo2, DeRoo1}. The analog result
for the closed superstring Gamma factor is given in (\ref{formula3}) and it is
quite remarkable that in this
case only $odd$ valued $\zeta(n)$ factors appear as coefficients of the
$\alpha'$ series.

\section*{Acknowledgements}

R. M. would like to thank N. Berkovits and F. Brandt for useful conversations
and the Brazilian agencies CNPq and FAPEMIG for partial financial support in
this research.  O. C. would like to thank J. Drummond and P. Vanhove for useful
comments,  FONDECYT grant 3000026 for partial
financial support and the Instituto Nazionale de Fisica Nucleare 
for a postdoctoral fellowship.

\appendix

\section{Conventions and Identities}
\label{conventions}

\begin{enumerate}
\item \underline{Metric}:
\begin{eqnarray}
\eta_{\mu \nu} = \mbox{diag}(-, +, \ldots , +).
\label{metric}
\end{eqnarray}
\item \underline{Spacetime indexes}:
\begin{eqnarray*}
\alpha, \beta & \rightarrow & \mbox{spinor components} \\
\kappa, \lambda, \mu, \nu, \rho, \sigma,  \tau, \phi, \omega &
\rightarrow & \mbox{vector components}
\end{eqnarray*}
\item \underline{Spinors and Dirac Matrices}:\\

Through out this paper $all$ spinors are 10-dimensional Majorana-Weyl,
with 16
real components.\\
The Dirac $\gamma^{\mu}$ matrices satisfy the Clifford algebra
\begin{eqnarray}
\label{anti-com}
\{ \gamma^{\mu}, \gamma^{\nu} \} = -2 \eta^{\mu \nu}.
\end{eqnarray}
Antisymmetric products
of $p$ $\gamma^{\mu}$ matrices, $\gamma_{\mu_1 \mu_2 \ldots \mu_p}$, are
defined with a global factor $1/p!$. For $\gamma_{\mu \nu \rho}$ we have
the
following identities:
\begin{eqnarray}
\label{gamma3}
\gamma_{\mu \nu \rho}= \gamma_{\mu}\gamma_{\nu}\gamma_{\rho}
+\gamma_{\mu}\eta_{\nu \rho} -\gamma_{\nu}\eta_{\rho \mu}
+\gamma_{\rho}\eta_{\mu \nu} \ ,
\end{eqnarray}
\begin{eqnarray}
\label{gamma4}
\gamma_{\mu \nu \rho}= \gamma_{\mu \nu}\gamma_{\rho} 
+\gamma_{\mu}\eta_{\nu \rho} -\gamma_{\nu}\eta_{\rho \mu} \ .
\end{eqnarray}
\noindent A useful identity for Majorana-Weyl spinors:
\begin{eqnarray}
\bar{\chi}_1 \gamma_{\mu_1 \ldots \mu_p} \chi_2 =
\pm (-1)^{p(p+1)/2}
\bar{\chi}_2 \gamma_{\mu_1 \ldots \mu_p} \chi_1 \ ,
\label{spinor3}
\end{eqnarray}
where the additional plus (minus) sign, before the
$(-1)^{p(p+1)/2}$ factor, is used for anticommuting
(commuting) spinors.\\

\noindent A Fierz identity:
\begin{eqnarray}
\bar{\chi}_1 \gamma_{\mu} \chi_2 \bar{\chi}_3 \gamma_{\mu} \chi_4 +
\bar{\chi}_1 \gamma_{\mu} \chi_3 \bar{\chi}_4 \gamma_{\mu} \chi_2 +
\bar{\chi}_1 \gamma_{\mu} \chi_4 \bar{\chi}_2 \gamma_{\mu} \chi_3 = 0 \ .
\label{Fierz}
\end{eqnarray}
In the way in which (\ref{Fierz}) has been written ,it is valid for
all Majorana-Weyl spinors $\chi_j$ being commuting as well as all of them
being
anticommuting.
\item \underline{Gauge group, generators and structure constants}:\\

Gauge fields are matrices in the $U(N)$ internal space, so that
$A_{\mu} = A^{\mu}_{\ a} \lambda^a$, where the hermitian generators
$\lambda^a$ are in the adjoint representation
\begin{eqnarray}
{(\lambda^a)}^{bc} = -i f^{abc \ }
\label{adjoint}
\end{eqnarray}
and they satisfy the usual relations
\begin{eqnarray}
[ \lambda^a, \lambda^b ] = i \ f^{ab}_{\ \ \ c} \ \lambda^c \ , \ \ \
\mbox{tr}(\lambda^a \lambda^b) = \delta^{ab} \ ,
\label{group1}
\end{eqnarray}
for real structure constants $f^{abc}$.\\
The field strength and the covariant derivative are defined by
\begin{eqnarray}
\label{group2}
F_{\mu \nu} & = & \partial_{\mu} A_{\nu} - \partial_{\nu} A_{\mu}
- ig [ A_{\mu}, A_{\nu}] \ , \\
\label{group3}
D_{\mu} \phi & = & \partial_{\mu} \phi - ig [ A_{\mu}, \phi ] \ ,
\end{eqnarray}
and they are related by the identity
\begin{eqnarray}
[D_{\mu}, D_{\nu}] \phi & = & -ig \ [ F_{\mu \nu}, \phi ] \ .
\label{group4}
\end{eqnarray}
Covariant derivatives of field strengths satisfy the Bianchi identity:
\begin{eqnarray}
\label{Bianchi}
D^{\mu} F^{\nu \rho} + D^{\rho} F^{\mu \nu} + D^{\nu} F^{\rho \mu} = 0 \ .
\end{eqnarray}
The following expressions naturally appear in the scattering amplitude
(\ref{A4}): \begin{eqnarray}
T_1 =  \tr(\lambda^{a_1} \lambda^{a_2} \lambda^{a_3} \lambda^{a_4}) +
\tr(\lambda^{a_4} \lambda^{a_3} \lambda^{a_2} \lambda^{a_1}) \ ,
\nonumber \\
T_2 = \mbox{tr}(\lambda^{a_1} \lambda^{a_3} \lambda^{a_2}
\lambda^{a_4}) +
\mbox{tr}(\lambda^{a_4} \lambda^{a_2} \lambda^{a_3} \lambda^{a_1}) \ ,
\nonumber \\
T_3 = \tr(\lambda^{a_1} \lambda^{a_2}
\lambda^{a_4} \lambda^{a_3}) +
\tr(\lambda^{a_3} \lambda^{a_4} \lambda^{a_2} \lambda^{a_1}) \ .
\label{T}
\end{eqnarray}
For them
\begin{eqnarray}
T_2-T_1 = - f^{a_1 a_4 b} f^{a_2 a_3}_{\ \ \ \ \ b} \ , \ \ \
T_3-T_2 = + f^{a_1 a_3 b} f^{a_2 a_4}_{\ \ \ \ \ b} \ , \ \ \
T_1-T_3 = - f^{a_1 a_2 b} f^{a_3 a_4}_{\ \ \ \ \ b} \ . \ \ \
\label{DeltaT}
\end{eqnarray}
Summing all three relations
comes the Jacobi identity:
\begin{eqnarray}
f^{a_1 a_4 b} f^{a_2 a_3}_{\ \ \ \ \ b} \
\ - \ f^{a_1 a_3 b} f^{a_2 a_4}_{\ \ \ \ \ b} \
\ + \ f^{a_1 a_2 b} f^{a_3 a_4}_{\ \ \ \ \ b} \ = \ 0 \ .
\label{Jacobi}
\end{eqnarray}
\item \underline{Identity for the $t_{(8)}$ tensor}:\\

The $t_{(8)}$ tensor\footnote{An explicit expression for it may be found
in
equation (4.A.21) of \cite{Schwarz}.}, characteristic of the 4 boson
scattering
amplitude,
is antisymmetric on each pair
$\mu_j \nu_j$ ($j=1,2,3,4$) and is symmetric under any exchange of such of
pairs.   It allows to write
the identity
\begin{multline}
P( 1, 2, 3, 4)  \ \
t^{(8)}_{\mu_1 \nu_1 \mu_2 \nu_2 \mu_3 \nu_3 \mu_4 \nu_4}
A^{\mu_1 \nu_1}_1 A^{\mu_2 \nu_2}_2
A^{\mu_3 \nu_3}_3 A^{\mu_4 \nu_4}_4 = \\
\begin{split}
& P( 1, 2, 3, 4)  \
\left( A_{1 \mu \nu} A_2^{\nu \rho}
A_{3  \rho \sigma}A_4^{\sigma \mu}
- \frac{1}{4} \ A_{1 \mu \nu}A_{2}^{\mu \nu}
A_{3 \rho \sigma}A_{4}^{\rho \sigma} \right) \ ,
\end{split}
\label{t8}
\end{multline}
for any $symmetric$ operator $P(1, 2, 3, 4)$, that acts on index $j$
of an antisymmetric tensor $A_j^{\mu \nu}$ ($j = 1,2,3,4$).

\end{enumerate}

\section{Derivation of the 4-particle scattering amplitudes from the
effective
action}
\label{derivation}

In this Appendix we follow closely the steps done for photons in the
corresponding Appendix of \cite{DeRoo1}. But before going into the
details,
we comment on the 4-particle
scattering processes that we are considering (at tree level).
It happens that
they always present 1-particle reducible and irreducible diagrams,
as the ones  shown
in (a), in figure \ref{four}. The 1-particle reducible diagrams arise only
in
the  zero order $\alpha'$ contribution of the effective theory, that is,
in the
Super Yang-Mills term ($S_{\rm SYM}$).\\
Since the method followed to derive the abelian formula
(\ref{Seff-abelian}) is based in the effective action as the generator of
1-particle irreducible diagrams only, we have excluded the zero
order $\alpha'$ contribution in (\ref{f}) and (\ref{Seff-nonabelian}) and
treated it separately in the $S_{\rm SYM}$ action of (\ref{sum}).
So, when deriving the 4-point scattering amplitudes in section
\ref{scattering} of this Appendix, we will calculate them as
\begin{eqnarray}
{\cal A}^{(4)} \ = \ {\cal A}_{(0)}^{(4)} \ + \
{\cal A}_{\rm \alpha' corr.}^{(4)} \ ,
\label{splitting}
\end{eqnarray}
where the zero order and the
$\alpha'$ contribution, ${\cal A}_{(0)}^{(4)}$ and
${\cal A}_{\rm \alpha' corr.}^{(4)}$, will be calculated
from the $S_{\rm SYM}$ and $S_{\rm \alpha' corr.}^{\ nonabel.}$ actions,
respectively.

\FIGURE[t]{\centerline{\epsfig{file=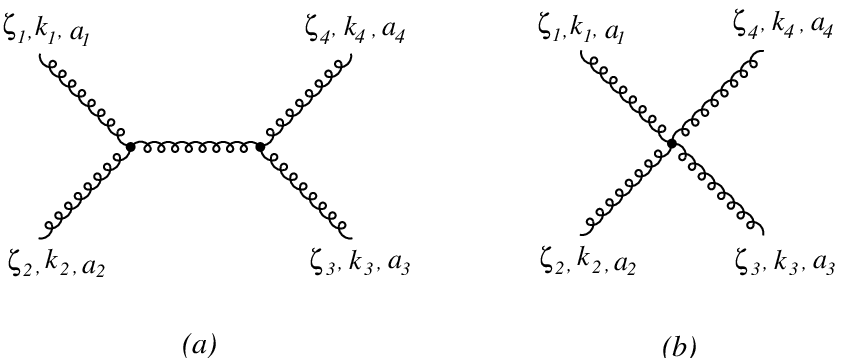}\caption{Type of tree
diagrams which contribute to the 4-boson scattering
amplitude in Super Yang-Mills theory. The diagram in (a) is 1-particle
reducible, while the diagram in (b) is 1-particle
irreducible.}\label{four}}}

\subsection{4-particle scattering amplitudes in SYM}
\label{SYM}

The (on-shell) tree level 4-point amplitudes in Super Yang-Mills
theory can be directly obtained from the $\alpha' \rightarrow 0$ limit of
the
corresponding amplitudes in Open Superstring theory. They are all given by
the common expression
\begin{eqnarray}
{\cal A}_{(0)}^{(4)} = 8 \ i \ g^2 \
(2 \pi)^{10} \delta^{(10)}(k_1 + k_2 + k_3 + k_4) \
f^{a_1 a_2 a_3 a_4}_{\rm (0)} (k_1, k_2, k_3, k_4) \
K^{(4)} \ ,
\label{compact1}
\end{eqnarray}
where
\begin{eqnarray}
f^{a_1 a_2 a_3 a_4}_{\rm (0)} (k_1, k_2, k_3, k_4) & = &
\{ \mbox{tr}(\lambda^{a_1} \lambda^{a_2} \lambda^{a_3} \lambda^{a_4}) +
\mbox{tr}(\lambda^{a_4} \lambda^{a_3} \lambda^{a_2} \lambda^{a_1}) \} \
\cdot \frac{1}{st} \nonumber \\
 & & + \{ \mbox{tr}(\lambda^{a_1} \lambda^{a_3} \lambda^{a_2}
\lambda^{a_4}) +
\mbox{tr}(\lambda^{a_4} \lambda^{a_2} \lambda^{a_3} \lambda^{a_1}) \} \
\cdot \frac{1}{tu} \nonumber \\
 & & + \{ \mbox{tr}(\lambda^{a_1} \lambda^{a_2}
\lambda^{a_4} \lambda^{a_3}) +
\mbox{tr}(\lambda^{a_3} \lambda^{a_4} \lambda^{a_2} \lambda^{a_1}) \} \
\cdot \frac{1}{us}  \ ,
\label{f1234}
\end{eqnarray}
and where $K^{(4)}$ denotes the kinematic factor corresponding to each
scattering process:
\begin{enumerate}
\item $4 \ bosons$:
\begin{eqnarray}
K_{\rm 4b} =
t_{(8)}^{\mu_1 \nu_1 \mu_2 \nu_2 \mu_3 \nu_3 \mu_4 \nu_4} \
\zeta^1_{\mu_1} k^1_{\nu_1} \zeta^2_{\mu_2} k^2_{\nu_2}
\zeta^3_{\mu_3} k^3_{\nu_3} \zeta^4_{\mu_4} k^4_{\nu_4} \ ,
\label{Kbosons}
\end{eqnarray}
where $\zeta_j$ and $k_j$ denote the $j$-th boson polarization and
momentum,
respectively. \\
$t_{(8)}^{\mu_1 \nu_1 \mu_2 \nu_2 \mu_3 \nu_3 \mu_4 \nu_4}$
is a completely
known tensor which satisfies, for example, the identity (\ref{t8}).

\item $2 \ bosons \ and \ 2 \ fermions$:
\begin{eqnarray}
K_{\rm 2b/2f} =  \frac{1}{8} \left[ \frac{}{}
(\bar{u}_1 {\zetaslash}_3 u_2 ) \left\{ \frac{}{} (k_2-k_1) \cdot \zeta_4
\
(k_3 \cdot k_4) - (k_2-k_1)\cdot k_4 \ (k_3 \cdot \zeta_4) \frac{}{}
\right\}
\right. \nonumber \\
+ \
(\bar{u}_1 {\kslash}_3 u_2 ) \left\{ \frac{}{} (k_2-k_1) \cdot k_4 \
(\zeta_3 \cdot \zeta_4) - (k_2-k_1)\cdot \zeta_4 \ (k_4 \cdot \zeta_3)
\frac{}{} \right\}  \nonumber \\
+ \
(\bar{u}_1 {\zetaslash}_4 u_2 ) \left\{ \frac{}{} (k_2-k_1) \cdot \zeta_3
\
(k_3 \cdot k_4) - (k_2-k_1)\cdot k_3 (k_4 \cdot \zeta_3) \frac{}{}
\right\}
\nonumber \\
+ \
(\bar{u}_1 {\kslash}_4 u_2 ) \left\{ \frac{}{} (k_2-k_1) \cdot k_3 \
(\zeta_3 \cdot \zeta_4) - (k_2-k_1)\cdot \zeta_3 \ (k_3 \cdot \zeta_4)
\frac{}{} \right\}  \nonumber \\
+ \
(\bar{u}_1 \gamma_{\mu \nu \rho} u_2) \left \{ \frac{}{}
(k_2 + k_1) \cdot \zeta_4 \
k_3^{\mu} k_4^{\nu} \zeta_3^{\rho}
- \
(k_2 + k_1) \cdot \zeta_3 \
k_3^{\mu} k_4^{\nu} \zeta_4^{\rho} \right. \nonumber \\
+ \left. \left. \
(k_2 + k_1) \cdot k_4 \
k_3^{\mu} \zeta_3^{\nu} \zeta_4^{\rho}
- \
(k_2 + k_1) \cdot k_3 \
k_4^{\mu} \zeta_3^{\nu} \zeta_4^{\rho} \frac{}{} \right\}
\frac{}{} \right] \ ,
\label{Kbosfer}
\end{eqnarray}
where $u_1$ and $u_2$ denote the fermion wave functions, $\zeta_3$ and
$\zeta_4$ denote the boson polarizations and $k_j$ denotes the $j$-th
particle
momentum.
\item $4 \ fermions$:
\begin{eqnarray}
K_{\rm 4f} =\frac{1}{24} \left[ \frac{}{} \left\{
k_1 \cdot k_4 + k_2 \cdot k_3 - k_1 \cdot k_3 - k_2 \cdot k_4
\right\} \ \bar{u}_1 \gamma_{\mu} u_2 \bar{u}_3
\gamma^{\mu} u_4 \right. \nonumber \\
+ \frac{}{} \left\{
k_1 \cdot k_2 + k_3 \cdot k_4 - k_1 \cdot k_4 - k_2 \cdot k_3
\right\} \ \bar{u}_1 \gamma_{\mu} u_3 \bar{u}_2
\gamma^{\mu} u_4 \nonumber \\
\left. +
\frac{}{} \left\{
k_1 \cdot k_3 + k_2 \cdot k_4 - k_1 \cdot k_2 - k_3 \cdot k_4
\right\} \ \bar{u}_1 \gamma_{\mu} u_4 \bar{u}_2
\gamma^{\mu} u_3
\frac{}{} \right] \ ,
\label{Kfermions}
\end{eqnarray}
where $u_j$ and $k_j$ denote the $j$-th fermion wave function and
momentum,
respectively.
\end{enumerate}
All these kinematic factors have alternative expressions when considering
the
momentum conservation and the on-shell conditions. For example,
using the $\gamma^{\mu}$ matrix relations in (\ref{anti-com}) and
(\ref{gamma3}), relation (\ref{spinor3}) for commuting spinors (with $p=1$
and
$p=3$),
the on-shell conditions  $\ k_3 \cdot \zeta_3 = k_4 \cdot \zeta_4 =
\bar{u}_2
\kslash_2 =\kslash_1 u_1 = \kslash_4 \kslash_4 = 0$, momentum conservation
and
introducing the Mandelstam variables (\ref{Mandelstam}), the kinematic
factors
in (\ref{Kbosfer}) and (\ref{Kfermions}) may be shortly written as
\begin{eqnarray}
K_{\rm 2b/2f} & = & -\frac{s}{8} \left[
\frac{}{} \bar{u}_2 \zetaslash_3 (\kslash_4 + \kslash_1)
\zetaslash_4 u_1 \frac{}{} \right] \nonumber \\
 & & -\frac{t}{8} \left[
\frac{}{} 2 \ (\bar{u}_2 \zetaslash_3 u_1) (k_3 \cdot \zeta_4)
+ 2 \ (\bar{u}_2 \kslash_4 u_1) (\zeta_3 \cdot \zeta_4)
- 2 \ (\bar{u}_2 \zetaslash_4 u_1) (k_4 \cdot \zeta_3)
 \frac{}{} \right]
\label{Kbosfer2}
\end{eqnarray}
and
\begin{eqnarray}
K_{\rm 4f} = -\frac{s}{8} \
\bar{u}_1 \gamma_{\mu} u_4 \bar{u}_2 \gamma^{\mu} u_3 \
+ \ \frac{t}{8} \
\bar{u}_1 \gamma_{\mu} u_2 \bar{u}_4 \gamma^{\mu} u_3 \ ,
\label{Kfermions2}
\end{eqnarray}
in agreement with \cite{Schwarz}, after taking into account the
observations of
\cite{Bergshoeff1} about the global factors in the 4-point
amplitudes\footnote{Due to the different conventions in the Mandelstam
variables, our formulas (\ref{Kbosfer2}) and (\ref{Kfermions2}) do not
look
directly the same as the ones in formulas (2.6)
and (2.4) of \cite{Bergshoeff1}, but the explicit dependence of the
kinematic
factors $in \ terms$ of the momenta $k_j$ is the same.}.\\
The way we have chosen to write the kinematic factors in (\ref{Kbosons}),
(\ref{Kbosfer}) and (\ref{Kfermions}) is such that the corresponding boson
and/or
fermion symmetry is manifest:  $K_{\rm 4b}$ is completely $symmetric$
under any particle interchange; $K_{\rm 2b/2f}$ is $antisymmetric$ under
interchanges of particles $1$ and $2$ (the fermions) and is $symmetric$
under
interchanges of particles $3$ and $4$ (the bosons); $K_{\rm 4f}$ is
completely $antisymmetric$ under any particle interchange.

\subsection{The 4-point effective action}
\label{4-pointaction}

In this Appendix we deal only with the $\alpha'$ correction terms
of the effective action, that is, $S_{\rm \alpha' corr.}^{\ nonabel.}$.
We begin writing it as the sum of two contributions:
\begin{eqnarray}
S_{\rm \alpha' corr.}^{\ nonabel.} [A, \psi ] =
S_{\rm \alpha' corr.}^{(4)} [A, \psi]
+ \ S_{\rm \alpha' corr.}^{(n > 4)} [A, \psi] \ ,
\label{decomposition}
\end{eqnarray}
where $S_{\rm \alpha' corr.}^{(4)} [A, \psi]$ contains only the quartic
field terms and $S_{\rm \alpha' corr.}^{(n > 4)} [A, \psi]$ contains the
higher field
terms (which are not present in the abelian theory).\\
The quartic field terms action in (\ref{decomposition}) may be written as
\begin{eqnarray}
S_{\rm \alpha' corr.}^{(4)} [A, \psi] & = & \int \int \int
\int  \left\{ \prod_{j=1}^4 d^{10}x_j \right \} \left[ \frac{}{}
\frac{1}{4!}
\Gamma_{\mu_1 \mu_2 \mu_3 \mu_4}^{(4) a_1 a_2 a_3 a_4}(x_1, x_2, x_3, x_4)
\
A_{a_1}^{\mu_1}(x_1) A_{a_2}^{\mu_2}(x_2) A_{a_3}^{\mu_3}(x_3)
A_{a_4}^{\mu_4}(x_4) \right. \nonumber \\
 & & + \ \frac{1}{2! \cdot 2!}
\Gamma_{\alpha_1 \alpha_2 \mu_3 \mu_4}^{(4) a_1 a_2 a_3 a_4}
(x_1, x_2, x_3, x_4) \ \psi_{a_1}^{\alpha_1}(x_1)
\psi_{a_2}^{\alpha_2}(x_2)
A_{a_3}^{\mu_3}(x_3) A_{a_4}^{\mu_4}(x_4) \nonumber \\
 & & \left. + \ \frac{1}{4!}
\Gamma_{\alpha_1 \alpha_2 \alpha_3 \alpha_4}^{(4) a_1 a_2 a_3 a_4} \
(x_1, x_2, x_3, x_4) \ \psi_{a_1}^{\alpha_1}(x_1)
\psi_{a_2}^{\alpha_2}(x_2)
\psi_{a_3}^{\alpha_3}(x_3) \psi_{a_4}^{\alpha_4}(x_4) \frac{}{} \right]
\label{4point}
\end{eqnarray}
where
\begin{eqnarray}
\Gamma_{\mu_1 \mu_2 \mu_3 \mu_4}^{(4) a_1 a_2 a_3 a_4}(x_1, x_2, x_3, x_4)
=
\left. \frac{\delta^4 S_{\rm \alpha' corr.}^{\ nonabel.} [A, \psi]}
{\delta A_{a_1}^{\mu_1}(x_1) \delta A_{a_2}^{\mu_2}(x_2)
\delta A_{a_3}^{\mu_3}(x_3) \delta A_{a_4}^{\mu_4}(x_4)}
\right|_{A_{a}^{\mu}=0, \ \psi_{a}^{\alpha}=0} \ ,  \nonumber \\
\Gamma_{\alpha_1 \alpha_2 \mu_3 \mu_4}^{(4) a_1 a_2 a_3 a_4}
(x_1, x_2, x_3, x_4) =
\left. \frac{\delta^4 S_{\rm \alpha' corr.}^{\ nonabel.} [A, \psi]}
{\delta \psi_{a_1}^{\alpha_1}(x_1) \delta \psi_{a_2}^{\alpha_2}(x_2)
\delta A_{a_3}^{\mu_3}(x_3) \delta A_{a_4}^{\mu_4}(x_4)}
\right|_{A_{a}^{\mu}=0, \ \psi_{a}^{\alpha}=0} \ ,  \nonumber \\
\Gamma_{\alpha_1 \alpha_2 \alpha_3 \alpha_4}^{(4) a_1 a_2 a_3 a_4} \
(x_1, x_2, x_3, x_4) =
\left. \frac{\delta^4 S_{\rm \alpha' corr.}^{\ nonabel.} [A, \psi]}
{\delta \psi_{a_1}^{\alpha_1}(x_1) \delta \psi_{a_2}^{\alpha_2}(x_2)
\delta \psi_{a_3}^{\alpha_3}(x_3) \delta \psi_{a_4}^{\alpha_4}(x_4)}
\right|_{A_{a}^{\mu}=0, \ \psi_{a}^{\alpha}=0} \ .
\label{proper}
\end{eqnarray}
The relations in (\ref{proper}) assume that the fermionic functional
derivatives are being taken from the right hand-side at any moment.\\
Using the explicit expression of $S_{\rm \alpha' corr.}^{\ nonabel.} [A,
\psi]$
in (\ref{Seff-nonabelian}), and the identity (\ref{t8}) for the $t_{(8)}$
tensor,
we obtain\footnote{In all these formulas we use the
notation $\partial_j^{\nu} = \partial / \partial x^j_{\nu}$.}:
\begin{multline}
\Gamma_{\mu_1 \mu_2 \mu_3 \mu_4}^{(4) a_1 a_2 a_3 a_4}(x_1, x_2, x_3, x_4)
 = \\
\begin{split}
&  - 8 \ g^2 \ {\alpha'}^2 \int d^{10}x \ \
t^{(8)}_{\mu_1 \nu_1 \mu_2 \nu_2 \mu_3 \nu_3 \mu_4 \nu_4}  \\
& {\cal G}^{\ a_1 a_2 a_3 a_4}_{\rm \alpha' corr.}
(\partial_1 \cdot \partial_2 +
\partial_3 \cdot \partial_4 \ , \partial_1 \cdot \partial_4 +
\partial_2 \cdot \partial_3 \ , \partial_1 \cdot \partial_3 +
\partial_2 \cdot \partial_4) \\
& \ \ \partial_1^{\nu_1} \delta^{(10)}(x-x_1)\partial_2^{\nu_2}
\delta^{(10)}(x-x_2) \partial_3^{\nu_3}
\delta^{(10)}(x-x_3)\partial_4^{\nu_4}
\delta^{(10)}(x-x_4) \ ,
\label{4-pointbosonic}
\end{split}
\end{multline}
\begin{multline}
\Gamma_{\alpha_1 \alpha_2 \mu_3 \mu_4}^{(4) a_1 a_2 a_3 a_4}
(x_1, x_2, x_3, x_4) = \\
\begin{split}
&  i  \ g^2 \ {\alpha'}^2 \int d^{10}x \ \
{\cal G}^{\ a_1 a_2 a_3 a_4}_{\rm \alpha' corr.}
(\partial_1 \cdot \partial_2 +
\partial_3 \cdot \partial_4 \ , \partial_1 \cdot \partial_4 +
\partial_2 \cdot \partial_3 \ , \partial_1 \cdot \partial_3 +
\partial_2 \cdot \partial_4) \\
& \ \ \left[ - \frac{}{} \
\left \{ \frac{}{} (\gamma^0 \gamma_{\mu_3})_{\alpha_1 \alpha_2} \
\partial^2_{\mu_4} \ \partial^3_{\nu} \ \partial_4^{\nu}
+
(\gamma^0 \gamma_{\mu_3})_{\alpha_1 \alpha_2} \
\partial^2_{\nu} \ \partial^3_{\mu_4} \ \partial_4^{\nu} \right. \right.\\
& \ \ \ \ \ \ \ \ + \left.
(\gamma^0 \gamma^{\nu})_{\alpha_1 \alpha_2} \  \eta_{\mu_3 \mu_4} \
\partial^2_{\rho} \ \partial^3_{\nu} \ \partial_4^{\rho}
+
(\gamma^0 \gamma^{\nu})_{\alpha_1 \alpha_2} \
\partial^2_{\mu_4} \ \partial^3_{\nu} \ \partial^4_{\mu_3} \frac{}{}
\right\}
\\
& \ \ \ + \
\left \{ \frac{}{} (\gamma^0 \gamma_{\sigma \mu_3 \rho})_{\alpha_1
\alpha_2}
\  \partial^2_{\mu_4} \ \partial_3^{\sigma} \ \partial_4^{\rho}
-
(\gamma^0 \gamma_{\sigma \mu_3 \mu_4})_{\alpha_1 \alpha_2} \
\partial_2^{\rho} \ \partial_3^{\sigma} \ \partial^4_{\rho}
\frac{}{} \right\} \\
& \ \ \ + \left.
\left( \begin{array}{c}
\mbox{18 terms obtained by antisymmetrizing
on indexes (1,2) and}\\
\mbox{symmetrizing on indexes (3,4)
the two previous curly brackets  }
\end{array} \right) \
\right] \\
& \ \ \ \ \delta^{(10)}(x-x_1) \delta^{(10)}(x-x_2)
\delta^{(10)}(x-x_3) \delta^{(10)}(x-x_4) \ ,
\label{4-pointbosfer}
\end{split}
\end{multline}
\begin{multline}
\Gamma_{\alpha_1 \alpha_2 \alpha_3 \alpha_4}^{(4) a_1 a_2 a_3 a_4}
(x_1, x_2, x_3, x_4)   = \\
\begin{split}
&   \frac{1}{6} \ g^2 \ {\alpha'}^2 \int d^{10}x \ \
{\cal G}^{\ a_1 a_2 a_3 a_4}_{\rm \alpha' corr.}
(\partial_1 \cdot \partial_2 +
\partial_3 \cdot \partial_4 \ , \partial_1 \cdot \partial_4 +
\partial_2 \cdot \partial_3 \ , \partial_1 \cdot \partial_3 +
\partial_2 \cdot \partial_4) \\
& \ \ \left[ \frac{}{} \
(\gamma^0 \gamma^{\mu})_{\alpha_1 \alpha_2}
(\gamma^0 \gamma_{\mu})_{\alpha_3 \alpha_4} \
\partial_2^{\nu} \partial^4_{\nu} \right. \\
& \ \ + \left. \left( \frac{}{}
\mbox{23 terms obtained by antisymmetrizing on indexes (1,2,3,4)}
\frac{}{} \right) \frac{}{} \  \right] \\
& \ \ \ \ \delta^{(10)}(x-x_1) \delta^{(10)}(x-x_2)
\delta^{(10)}(x-x_3) \delta^{(10)}(x-x_4) \ ,
\label{4-pointfermionic}
\end{split}
\end{multline}
where ${\cal G}^{\ a_1 a_2 a_3 a_4}_{\rm \alpha' corr.}
(s , t , u)$ is given in (\ref{Gnonabelian}).\\
By definition (see (\ref{proper})), all 4-point functions have the
appropriate boson and/or fermion symmetry, as was commented in the
previous
subsection in the case of the kinematic factors.
This boson and/or fermion symmetry can be checked directly in the
expressions (\ref{4-pointbosonic}), (\ref{4-pointbosfer}) and
(\ref{4-pointfermionic}).

\subsection{The scattering amplitudes}
\label{scattering}

>From the previous 4-point functions, the corresponding 4-point scattering
amplitudes are calculated, in our conventions, as
\begin{eqnarray}
{\cal A}_{\rm \alpha' corr.}^{(4)} = -i \ (2 \pi)^{10}
\delta^{(10)}(k_1 + k_2 + k_3 + k_4) \
\phi_1^{p_1} \phi_2^{p_2} \phi_3^{p_3} \phi_4^{p_4} \
\Gamma_{p_1 p_2 p_3 p_4}^{(4) a_1 a_2 a_3 a_4}
(k_1, k_2, k_3, k_4) \ ,
\label{complete}
\end{eqnarray}
where
\begin{multline}
(2 \pi)^{10} \delta^{(10)}(k_1 + k_2 + k_3 + k_4) \
\Gamma_{p_1 p_2 p_3 p_4}^{(4) a_1 a_2 a_3 a_4}(k_1, k_2, k_3, k_4) = \\
\begin{split}
& \int \int \int \int \left\{ \prod_{j=1}^4 d^{10}x_j \
e^{i k_j \cdot x_j} \right\}
\Gamma_{p_1 p_2 p_3 p_4}^{(4) a_1 a_2 a_3 a_4}
(x_1, x_2, x_3, x_4) \ .
\label{Fourier}
\end{split}
\end{multline}
In (\ref{complete}), $p_j$ and $\phi_j^{p_j}$ denote, respectively, the
$j$-th
vector (spinor) index and the $j$-th boson polarization vector (fermion
wave
function), according to the specific scattering process.\\
Upon substituting (\ref{4-pointbosonic}), (\ref{4-pointbosfer}) and
(\ref{4-pointfermionic}) in (\ref{Fourier}) and then in (\ref{complete}),
all three cases can be summarized as
\begin{eqnarray}
{\cal A}_{\rm \alpha' corr.}^{(4)} = 8 \ i \ g^2 \ {\alpha'}^2
(2 \pi)^{10} \delta^{(10)}(k_1 + k_2 + k_3 + k_4) \
{\cal G}^{\ a_1 a_2 a_3 a_4}_{\rm \alpha' corr.}
(s , t , u) \ K^{(4)} \ ,
\label{compact2}
\end{eqnarray}
where, for each scattering process, $K^{(4)}$ is exactly the same
kinematic
factor given in (\ref{Kbosons}), (\ref{Kbosfer}) and (\ref{Kfermions}).\\
So, substituting (\ref{compact1}) and (\ref{compact2}) in (\ref{splitting})
we
finally see that our nonabelian 4-point effective action in (\ref{sum}),
(\ref{SuperYangMills}) and (\ref{Seff-nonabelian}),
reproduces the known 4-particle
scattering amplitudes of Open Superstring theory:
\begin{eqnarray}
{\cal A}^{(4)} & = &
8 \ i \ g^2 {\alpha'}^2 (2 \pi)^{10}
\delta^{(10)}(k_1 + k_2 + k_3 +k_4)\times \nonumber \\&&{}\times  \Biggl[
\frac{\Gamma(- \alpha' s)  \Gamma(- \alpha' t)}{\Gamma(1- \alpha' s -
\alpha'
t)} \{ \tr(\lambda^{a_1} \lambda^{a_2} \lambda^{a_3} \lambda^{a_4}) +
\tr(\lambda^{a_4} \lambda^{a_3} \lambda^{a_2} \lambda^{a_1}) \}
 \nonumber \\ & &
\hphantom{{}\times  \Biggl[}
+\frac{\Gamma(- \alpha' t) \Gamma(- \alpha' u)}{\Gamma(1- \alpha' t -
\alpha' u)} \{ \tr(\lambda^{a_1} \lambda^{a_3} \lambda^{a_2}
\lambda^{a_4}) +
\tr(\lambda^{a_4} \lambda^{a_2} \lambda^{a_3} \lambda^{a_1}) \}
 \nonumber \\ & &
\hphantom{{}\times  \Biggl[}
 + \frac{\Gamma(- \alpha' u) \Gamma(- \alpha' s)}{\Gamma(1-
\alpha' u - \alpha' s)} \{ \tr(\lambda^{a_1} \lambda^{a_2}
\lambda^{a_4} \lambda^{a_3}) +
\tr(\lambda^{a_3} \lambda^{a_4} \lambda^{a_2} \lambda^{a_1}) \}
 \Biggr] \cdot K^{(4)} \,.
\label{compact3}
\end{eqnarray}

\section{$f \mbox{-} f$ part of the ${\cal O}({\alpha'}^5)$
action}
\label{S5}

The ${\cal S}_{(5), f \mbox{-} f}$ action in (\ref{S52}) is given by
\begin{multline}
{\cal S}_{(5), f \mbox{-} f}  =  \\
\begin{split}
&  \int d^{10}x \
f^{a_1 a_2 b} f^{a_3 a_4}_{\ \ \ \ \ b} \ \left\{ \frac{}{}
\left[ \frac{}{}
D_{\kappa} D_{\tau}
D_{\lambda} F_{a_1 \mu \nu}
D^{\kappa} D^{\tau}
F_{a_2}^{\nu \rho}
D^{\lambda} F_{a_3 \rho \sigma}
F_{a_4}^{\sigma \mu} \right. \right. \\
&  - \frac{1}{4} \
D_{\kappa} D_{\tau}
D_{\lambda}F_{a_1 \mu \nu}
D^{\kappa} D^{\tau}
F_{a_2}^{\mu \nu}
D^{\lambda}F_{a_3 \rho \sigma}
F_{a_4}^{\rho \sigma}
+ \ 2i \ D_{\kappa} D_{\tau}
D_{\lambda}\bar{\psi}_{a_1}
\gamma^{\nu} D^{\kappa} D^{\tau}
D_{\rho}\psi_{a_2}
D^{\lambda}F_{a_3 \nu \mu}
F_{a_4}^{\mu \rho} \\
&  - \ i \ D_{\kappa} D_{\tau}
D_{\lambda}\bar{\psi}_{a_1}
\gamma^{\mu \nu \rho} D^{\kappa} D^{\tau}
D^{\sigma} \psi_{a_2}
D^{\lambda} F_{a_3 \mu \nu}
F_{a_4 \rho \sigma} \\
&  \left.  -\frac{1}{3} \ D_{\kappa} D_{\tau}
D_{\lambda}\bar{\psi}_{a_1}\gamma_{\mu}
D^{\kappa} D^{\tau}
D_{\nu} \psi_{a_2}
D^{\lambda}\bar{\psi}_{a_3}\gamma^{\mu}
D^{\nu} \psi_{a_4} 
\frac{}{} \right] \\
&  + \ \left[ \frac{}{}
D_{\kappa} D_{\tau}
D_{\lambda} F_{\mu \nu \ a_1}
F^{\nu \rho}_{\ a_2}
D^{\lambda} F_{a_3 \rho \sigma}
D^{\kappa} D^{\tau}
F_{a_4}^{\sigma \mu} \right. \\
&  - \frac{1}{4} \
D_{\kappa} D_{\tau}
D_{\lambda}F_{a_1 \mu \nu}
F_{a_2}^{\mu \nu}
D^{\lambda}F_{a_3 \rho \sigma}
D^{\kappa} D^{\tau}
F_{a_4}^{\rho \sigma}
+ \ 2i \ D_{\kappa} D_{\tau}
D_{\lambda}\bar{\psi}_{a_1}
\gamma^{\nu}
D_{\rho}\psi_{a_2}
D^{\lambda}F_{a_3 \nu \mu}
D^{\kappa} D^{\tau}
F_{a_4}^{\mu \rho} \\
&  - \ i \ D_{\kappa} D_{\tau}
D_{\lambda}\bar{\psi}_{a_1}
\gamma^{\mu \nu \rho}
D^{\sigma} \psi_{a_2}
D^{\lambda} F_{a_3 \mu \nu}
D^{\kappa} D^{\tau}
F_{a_4 \rho \sigma} \\
&  \left. -\frac{1}{3} \ D_{\kappa} D_{\tau}
D_{\lambda}\bar{\psi}_{a_1}\gamma_{\mu}
D_{\nu} \psi_{a_2}
D^{\lambda}\bar{\psi}_{a_3}\gamma^{\mu}
D^{\kappa} D^{\tau}
D^{\nu} \psi_{a_4} 
\frac{}{} \right] \\ 
&  + \ \left[ \frac{}{}
D_{\kappa} D_{\tau}
D_{\lambda} F_{a_1 \mu \nu}
F_{a_2}^{\nu \rho}
D^{\kappa} D^{\tau}
D^{\lambda} F_{a_3 \rho \sigma}
F_{a_4}^{\sigma \mu} \right. \\
&  - \frac{1}{4} \
D_{\kappa} D_{\tau}
D_{\lambda}F_{a_1 \mu \nu}
F_{a_2}^{\mu \nu}
D^{\kappa} D^{\tau}
D^{\lambda}F_{a_3 \rho \sigma}
F_{a_4}^{\rho \sigma}
+ \ 2i \ D_{\kappa} D_{\tau}
D_{\lambda}\bar{\psi}_{a_1}
\gamma^{\nu}
D_{\rho}\psi_{a_2}
D^{\kappa} D^{\tau}
D^{\lambda}F_{a_3 \nu \mu}
F_{a_4}^{\mu \rho} \\
&  - \ i \ D_{\kappa} D_{\tau}
D_{\lambda}\bar{\psi}_{a_1}
\gamma^{\mu \nu \rho}
D^{\sigma} \psi_{a_2}
D^{\kappa} D^{\tau}
D^{\lambda} F_{a_3 \mu \nu}
F_{a_4 \rho \sigma} \\
&  \left. -\frac{1}{3} \ D_{\kappa} D_{\tau}
D_{\lambda}\bar{\psi}_{a_1}\gamma_{\mu}
D_{\nu} \psi_{a_2}
D^{\kappa} D^{\tau}
D^{\lambda}\bar{\psi}_{a_3}\gamma^{\mu}
D^{\nu} \psi_{a_4} 
\frac{}{} \right] \\ 
& +  \int d^{10}x \ 
f^{a_1 a_3 b} f^{a_2 a_4}_{\ \ \ \ \ b} \ 
\left\{ \frac{}{} \left[ \frac{}{} 
D_{\kappa} D_{\tau} 
D_{\lambda} F_{a_1 \mu \nu} 
D^{\kappa} D^{\tau} 
D^{\lambda} F_{a_2}^{\nu \rho} 
F_{a_3 \rho \sigma} 
F_{a_4}^{\sigma \mu} 
\right. \right. \\ 
&- \frac{1}{4} \ 
D_{\kappa} D_{\tau} 
D_{\lambda}F_{a_1 \mu \nu} 
D^{\kappa} D^{\tau} 
D^{\lambda} F_{a_2}^{\mu \nu} 
F_{a_3 \rho \sigma} 
F_{a_4}^{\rho \sigma} 
+ \ 2i \ 
D_{\kappa} D_{\tau} 
D_{\lambda}\bar{\psi}_{a_1} 
\gamma^{\nu} 
D^{\kappa} D^{\tau} 
D^{\lambda}D_{\rho}\psi_{a_2} 
F_{a_3 \nu \mu} 
F_{a_4}^{\mu \rho}  \\ 
&  - \ i \ 
D_{\kappa} D_{\tau} 
D_{\lambda}\bar{\psi}_{a_1} 
\gamma^{\mu \nu \rho} 
D^{\kappa} D^{\tau} 
D^{\lambda}D^{\sigma} \psi_{a_2} 
F_{a_3 \mu \nu} 
F_{a_4 \rho \sigma}  \\ 
&  \left. -\frac{1}{3} \ 
D_{\kappa} D_{\tau} 
D_{\lambda}\bar{\psi}_{a_1}\gamma_{\mu} 
D^{\kappa} D^{\tau} 
D^{\lambda}D_{\nu} \psi_{a_2} 
\bar{\psi}_{a_3}\gamma^{\mu} 
D^{\nu} \psi_{a_4} 
\frac{}{} \right] \\ 
& +  \left[ \frac{}{} 
D_{\kappa} D_{\tau} 
D_{\lambda} F_{a_1 \mu \nu} 
D^{\lambda} F_{a_2}^{\nu \rho} 
F_{a_3 \rho \sigma} 
D^{\kappa} D^{\tau} 
F_{a_4}^{\sigma \mu} 
\right. \\ 
&- \frac{1}{4} \ 
D_{\kappa} D_{\tau} 
D_{\lambda}F_{a_1 \mu \nu} 
D^{\lambda} F_{a_2}^{\mu \nu} 
F_{a_3 \rho \sigma} 
D^{\kappa} D^{\tau} 
F_{a_4}^{\rho \sigma} 
+ \ 2i \ 
D_{\kappa} D_{\tau} 
D_{\lambda}\bar{\psi}_{a_1} 
\gamma^{\nu} 
D^{\lambda}D_{\rho}\psi_{a_2} 
F_{a_3 \nu \mu} 
D^{\kappa} D^{\tau} 
F_{a_4}^{\mu \rho}  \\ 
&  - \ i \ 
D_{\kappa} D_{\tau} 
D_{\lambda}\bar{\psi}_{a_1} 
\gamma^{\mu \nu \rho} 
D^{\lambda}D^{\sigma} \psi_{a_2} 
F_{a_3 \mu \nu} 
D^{\kappa} D^{\tau} 
F_{a_4 \rho \sigma}  \\ 
&  \left. -\frac{1}{3} \ 
D_{\kappa} D_{\tau} 
D_{\lambda}\bar{\psi}_{a_1}\gamma_{\mu} 
D^{\lambda}D_{\nu} \psi_{a_2} 
\bar{\psi}_{a_3}\gamma^{\mu} 
D^{\kappa} D^{\tau} 
D^{\nu} \psi_{a_4}
\frac{}{} \right] \\  
& + \left[ \frac{}{} 
D_{\kappa} D_{\tau} 
D_{\lambda} F_{a_1 \mu \nu} 
D^{\lambda} F_{a_2}^{\nu \rho} 
D^{\kappa} D^{\tau} 
F_{a_3 \rho \sigma} 
F_{a_4}^{\sigma \mu} 
\right.  \\ 
&- \frac{1}{4} \ 
D_{\kappa} D_{\tau} 
D_{\lambda}F_{a_1 \mu \nu} 
D^{\lambda} F_{a_2}^{\mu \nu} 
D^{\kappa} D^{\tau} 
F_{a_3 \rho \sigma} 
F_{a_4}^{\rho \sigma} 
+ \ 2i \ 
D_{\kappa} D_{\tau} 
D_{\lambda}\bar{\psi}_{a_1} 
\gamma^{\nu} 
D^{\lambda}D_{\rho}\psi_{a_2} 
D^{\kappa} D^{\tau} 
F_{a_3 \nu \mu} 
F_{a_4}^{\mu \rho}  \\ 
&  - \ i \ 
D_{\kappa} D_{\tau} 
D_{\lambda}\bar{\psi}_{a_1} 
\gamma^{\mu \nu \rho} 
D^{\lambda}D^{\sigma} \psi_{a_2} 
D^{\kappa} D^{\tau} 
F_{a_3 \mu \nu} 
F_{a_4 \rho \sigma}  \\ 
&  \left.  -\frac{1}{3} \ 
D_{\kappa} D_{\tau} 
D_{\lambda}\bar{\psi}_{a_1}\gamma_{\mu} 
D^{\lambda}D_{\nu} \psi_{a_2} 
D^{\kappa} D^{\tau} 
\bar{\psi}_{a_3}\gamma^{\mu} 
D^{\nu} \psi_{a_4} 
\frac{}{} \right]   
\left. \frac{}{} \right\} 
\end{split}
\label{S51a}
\end{multline}
We have directly used eq. (\ref{S51}) to derive this expression, without 
looking for any simplification when 
using integration by parts, equations of motion or the identities in
Appendix
\ref{conventions}. So it may well happen that the number of terms in this
action
reduces if these last considerations are taken into account.

\end{document}